\documentclass[acmsmall,screen,nonacm]{acmart}
\acmJournal{TOCHI}
\AtBeginDocument{%
  }

\setcopyright{acmlicensed}
\copyrightyear{2026}
\acmYear{2026}
\acmDOI{XXXXXXX.XXXXXXX}

\begin{document}

\title{Designing Needs- and Attention-Aware AI Learning Tools for Engineering Education: Insights from Psychological Outcomes}

\author{Kevin Zhongyang Shao}
\affiliation{%
  \institution{University of Washington}
  \city{Seattle}
  \state{WA}
  \country{USA}
}
\email{kshao918@uw.edu}

\author{Denise Wilson}
\affiliation{%
  \institution{University of Washington}
  \city{Seattle}
  \state{WA}
  \country{USA}
}
\email{denisew@uw.edu}

\author{Yale Quan}
\affiliation{%
  \institution{University of Southern Mississippi}
  \city{Hattiesburg}
  \state{MS}
  \country{USA}
}
\email{Yale.Quan@usm.edu}

\author{Sep Makhsous}
\affiliation{%
  \institution{University of Washington}
  \city{Seattle}
  \state{WA}
  \country{USA}
}
\email{sosper30@uw.edu}

\renewcommand{\shortauthors}{Shao et al.}

\begin{abstract}
Artificial Intelligence (AI) is transforming higher education, but its benefits can vary depending on where, how, and how often it supports learning. While prior research emphasizes cognitive and academic outcomes, this study examines how AI chatbots support the psychological needs and motivational states of engineering students. A survey of college engineering students (n = 206) examined perceived effects of AI chatbots on autonomy, relatedness, and relief from competence frustration. Structural equation modeling with latent interaction effects examined how baseline autonomy, competence frustration, relatedness, and personal agency contributed to perceived AI outcomes. Results indicate that students perceived that AI provided the greatest benefits as relief from competence frustration, smaller benefits for autonomy, and the weakest benefits for relatedness. Baseline motivational states mattered more than demographic factors, and inattention moderated how baseline competence frustration and autonomy related to perceived AI-related benefits. These results offer insights into formulating design principles for engineering-specific AI-based tools.  
\end{abstract}


\received{5 July 2026}
\received[revised]{XXX}
\received[accepted]{XXX}

\maketitle

\section{Introduction}
Artificial intelligence (AI) tools - especially large-language-model (LLM) - based tutors and assistants—are increasingly embedded in higher education. They can deliver contingent, stepwise feedback, worked-to-faded examples, and flexible pacing that support complex problem solving \cite{vanlehn2011relative, atkinson2003transitioning, weinman2021improving, winget2022practical}. Recent classroom deployments and field studies show that LLM tutors can be integrated at scale and shape how students plan, request hints, and persist through challenges \cite{kazemitabaar2024codeaid, lieb2024student}. Despite rapid progress in computing and physics, purpose-built tutors for engineering subdisciplines (e.g., circuit analysis, signal processing) remain limited, with a small but growing set of prototypes and early deployments \cite{knievel2025aitee}.

In spite of these limitations, a growing body of research indicates small-to-moderate gains in cognitive outcomes (task performance, efficiency, personalization) for educational chatbots/LLM tutors \cite{labadze2023role, wu2024ai, wang2025effect}. By contrast, studies that focus on affective outcomes or outcomes that blend cognitive and affective elements have been comparatively sparse. Many of these studies are further limited by the fact that they rely on short-term perception measures rather than validated instruments and are largely cross-sectional rather than longitudinal in nature. Even recent EdTech deployments that instrument LLM tutors are only beginning to capture motivational processes in situ, leaving open questions about which psychological outcomes are most responsive to AI support, who benefits most, and under what conditions \cite{lo2024influence}.

While a wide range of non-cognitive outcomes are of interest in student learning, this study is grounded in the three fundamental psychological needs posited by self-determination theory (SDT). SDT identifies these fundamental needs as grounded in an individual’s sense of competence, autonomy, and relatedness. The fulfillment of these needs fosters high-quality motivation while failure to satisfy them leads to demotivation and struggles to persist \cite{ryan2020intrinsic, ryan2018self}. Related to competence, self-efficacy beliefs reflect one’s perceived capability in a specific domain and shapes effort, persistence, and strategy use \cite{schunk2020motivation, schunk2016self, basileo2024role}. Recent papers on LLM tutors and teachable-agent suggest promising design trends - conversational scaffolding that keeps goals salient, resisting “give-away” answers, and learning-by-teaching setups that may boost agency and competence - but systematic evidence on psychological change is still scarce \cite{jin2024teach, kazemitabaar2024codeaid, duan2022designing, luo2025design}.

While addressing psychological needs is important for all students, they matter acutely for students who struggle with sustained attention, including many with ADHD. For these learners, motivational and self-regulatory outcomes—basic psychological needs (autonomy, competence, relatedness), self-beliefs (self-efficacy), and self-regulated learning—are proximal drivers of initiation, persistence, and effective help-seeking \cite{ryan2018self, karabenick2011understanding, zimmerman2002becoming, bandura1997self}. ADHD frequently persists into adulthood; recent U.S. surveillance estimates ~6\% of adults have a current diagnosis, with under-treatment common \cite{staley2024attention}. Studies of college populations indicate prevalence of diagnosed and undiagnosed ADHD and highlight executive-function barriers (planning, working memory, time management) that can impede success \cite{scheres2021adhd, sedgwick2022university, alvarez2023systematic}. Recent studies have called for neurodiversity-affirming design in higher education and have begun exploring how GenAI can support daily academic tasks for disabled and neurodivergent students \cite{tcherdakoff2025designing, glazko2023autoethnographic, atcheson2025d}.

Because of the increased importance of psychological needs, self-efficacy, and motivation in facilitating the success or failure of students who struggle with maintaining appropriately focused attention, this paper examines AI chatbot use through the lens of students’ psychological needs in everyday engineering learning contexts, with particular attention to learner heterogeneity across the attention spectrum. In so doing, it sets the stage for bridging potential equity gaps for students who struggle with attention, both those gaps that existed before AI learning tools emerged in engineering education and those that have resulted from AI tool use. To achieve this goal, quantitative analysis of close-ended survey questions is used to explore relationships among demographics, psychological needs outcomes, attention challenges, and perceived benefits of chatbot use. Exploratory and confirmatory factor analyses are first used to evaluate the adapted measurement structure, followed by structural equation modeling and latent interaction analyses to examine associations among baseline learner characteristics and perceived AI-related outcomes. This approach enables the examination of the following research questions:
\begin{itemize}
\item \textbf{Research Question 1 (RQ1):} What is the impact of AI tool use on students’ psychological needs?
\item \textbf{Research Question 2 (RQ2):} How are these impacts affected by general satisfaction of psychological needs, self-regulated learning, and self-efficacy beliefs?
\item \textbf{Research Question 3 (RQ3):} How do students with attention-related challenges experience AI differently from other students?
\end{itemize}
This paper makes the following contributions:
\begin{itemize}
\item \textbf{Empirical characterization of need-specific AI-related outcomes.} Rather than treating AI benefit as uniform, this study shows that perceived benefits from AI chatbot use were strongest for relief from competence frustration, more moderate for autonomy, and weakest for relatedness.
\item \textbf{Evidence on heterogeneous effects in AI-mediated learning.} The study shows that baseline competence frustration, autonomy, personal agency, and inattention shaped perceived AI-related outcomes more strongly than demographic composition, and that inattention moderated how baseline competence frustration and autonomy translated into perceived benefits.
\item \textbf{Reusable HCI design and evaluation knowledge.} The study translates these findings into implications for designing needs- and attention- aware AI learning tools that employ mechanism-level measures, heterogeneous-effects analysis, and in-process telemetry to enable dynamic adaptation to student learning needs.
\end{itemize}

\section{Related Work}
Existing research has suggested that AI tools support the satisfaction of the core psychological needs of autonomy, competence, and relatedness as posited by self-determination theory (SDT) \cite{deci2000and, ryan2000self, deci2013intrinsic, ballou2022self}. These needs are central to sustaining motivation and well-being in higher education \cite{niemiec2009autonomy, neufeld2019exploring}, with competence reflecting students’ sense of mastery, autonomy capturing their agency in directing learning, and relatedness shaping their experience of connection and support \cite{ryan2000self}. In human-computer interaction (HCI), SDT has also been used as a design lens for interactive systems, including conversational agents, suggesting that technology design can support or undermine these needs through features such as choice, feedback, and social framing \cite{yang2021designing}. Recent HCI work further cautions that SDT should not be used only as a broad vocabulary for evaluating positive user experience, but should instead guide the specification of motivational mechanisms, design features, and boundary conditions \cite{tyack2024self}. AI interventions that effectively address these needs are positioned to promote deeper engagement, stronger learning outcomes, and more supportive user experiences \cite{ryan2000self, ballou2022self}. The effective use of AI tools does not influence these needs in isolation, however, with self-regulated learning and self-efficacy beliefs playing key roles in the impact of these tools.

\subsection{AI Support for Psychological Needs of Autonomy, Competence, and Relatedness}
Adaptive and personalized scaffolding -- tailoring prompts and feedback to a learner model in real time -- is a core mechanism of AI-supported learning. AI-based tutoring systems such as MetaTutor \cite{azevedo2022lessons} and Mentigo \cite{zha2025mentigo} embed prompts that scaffold learners’ planning, monitoring, and reflection; evidence from higher education and second-language learning contexts show such AI support can boost confidence and reduce anxiety \cite{du2024transforming, zhai2024effects}. Empirically, confidence is associated with greater persistence and performance, consistent with confidence catalyzing motivated action that, in turn, furnishes mastery experiences which support competence need satisfaction \cite{multon1991relation, white1959motivation, ryan2000self}. In parallel, AI-enabled recommendation systems improved students’ engagement and performance \cite{huang2023effects}. SDT research shows that providing meaningful choice and supporting self-endorsed goals functions as autonomy support and enhances motivation; thus, AI-based recommendation systems that afford real choice in pathways/goals are well-positioned to support autonomy and, in turn, motivation \cite{patall2008effects, jang2010engaging}. Reflection-oriented systems such as Muse prompt students to revisit and justify their reasoning, reinforcing metacognitive awareness and self-regulatory practice \cite{cabales2019muse}. Such prompts have been shown to increase perceived competence in chatbot-supported classrooms \cite{yin2025effects, yin2024using} and can enact cognitive autonomy support when they invite students to justify their solutions \cite{stefanou2004supporting}.

Conversational agents and large language model (LLM)-based companions provide another important avenue of support. From an HCI perspective, they are especially relevant because their interaction design can directly support or frustrate users’ needs for competence, autonomy, and relatedness. Prior HCI work has applied SDT to conversational agents and argues that non-controlling guidance, meaningful choice, and supportive social framing are key design aspects for need-supportive interaction \cite{yang2021designing}. Task-parsing systems like VAL reduce cognitive load by breaking complex problems into manageable steps, reinforcing learners’ confidence in their ability to succeed \cite{kirchner2024outplay}. Decomposing problems into steps can provide structure; \citet{jang2010engaging} states that structure together with autonomy support is what fosters engagement and competence. Chatbot-assisted writing systems align with competence-supportive effects (e.g., confidence gains and reduced anxiety) when feedback is informational rather than controlling \cite{du2024transforming, niemiec2009autonomy}. Structured dialog frameworks such as the Conversation Progress Guide increase users’ confidence (self-efficacy) during conversational AI tasks \cite{jeong2025conversation}, which could subsequently improve the competence. To the extent that conversational agents cultivate social presence, they can also support relatedness in learning settings \cite{hew2023using}; in classroom use, supportive pedagogy around chatbots provides further opportunity to serve relatedness needs \cite{chiu2024teacher}. More recent work on LLM companions, including ChatGPT integrated into modeling environments \cite{chen2024learning} and IDE-embedded coding support \cite{kazemitabaar2024codeaid}, suggests that conversational AI can support learning through turn-by-turn guidance, follow-up clarification, and iterative refinement of ideas. These conversational properties matter because they allow learners to ask for the next step, test partial ideas, request clarification, and adjust the level of support while remaining in control of the interaction. In this way, conversational AI can support competence by providing timely informational feedback and support autonomy by allowing learners to preserve user control over the interaction \cite{ryan2000self, yang2021designing}.

Immersive technologies such as virtual reality (VR) and augmented reality (AR) also have strong potential to support psychological needs through embodied interaction and informational feedback and guidance. In VR environments, need-satisfaction evidence shows that design features that increase natural mapping or support user control raise autonomy and competence satisfaction, which in turn improve intrinsic motivation and engagement \cite{reer2022virtual, ijaz2020player}. In educational VR storytelling, participants reported a clear sense of autonomy from being able to explore the scene, but competence and relatedness lagged when interactivity and context were limited – pinpointing design levers for supporting all three needs \cite{green2024understanding}. In AR, a handheld anatomy game in midwifery education was associated with significant increases in Intrinsic Motivation Inventory (IMI)-measured perceived competence and intrinsic motivation after use, illustrating the potential of interactive AR feedback and practice opportunities to support competence-related motivation \cite{blattgerste2022motivational}. Meanwhile, in AR classrooms, an SDT-guided educational escape activity was explicitly designed to support autonomy (through limited difficulty tasks accompanied by multiple options for those tasks), competence (via appropriately challenging tasks with informational feedback), and relatedness (facilitating team collaboration). In a randomized comparison, both the AR and non-AR conditions showed significant within-group learning gains, but the gains did not differ significantly between groups, nor were there significant between-group differences in intrinsic motivation or post-test scores \cite{elford2022fostering}. 

While these studies underscore the potential of AI tools to support autonomy, competence, and relatedness, much of the existing evidence comes from structured or task-specific learning contexts in which AI use is researcher- or instructor-guided. Less is known about how these tools function in students’ everyday higher-education use across courses and self-directed academic tasks. This study contributes to the existing literature by examining how engineering students in higher education in the United States with different baseline psychological needs experience AI tools in real, everyday academic contexts. 

\subsection{The Influence of Self-Regulated Learning and Self-Efficacy Beliefs on AI Support for Psychological Needs}
Self-regulated learning (SRL) refers to learners’ cyclical processes of forethought, monitoring, strategy control, and reflection that coordinate cognition, motivation, and behavior toward goals \cite{zimmerman2002becoming, pintrich2004conceptual}. Ample evidence across educational settings suggests that SRL influences the connection between psychological needs satisfaction and academic outcomes. SRL- oriented scaffolds such as MetaTutor’s planning/monitoring prompts \cite{azevedo2022lessons}, Help Tutor’s feedback on help-seeking \cite{aleven2016help}, AutoTutor’s adaptive dialogue \cite{graesser2016conversations, nye2014autotutor} improve regulation behaviors and yield learning gains. Meta-analytic evidence further ties SRL strategy use to achievement (e.g., GPA/grades), and SRL intervention meta-analyses show concurrent gains in motivational outcomes (e.g., self-efficacy, motivation), tying SRL to both cognitive and affective results \cite{theobald2021self, dignath2008can, chen2022effectiveness, richardson2012psychological, crede2011meta}. Grounded in SDT, SRL scaffolds provide structure and effectance-relevant feedback (supporting competence) and meaningful choice (supporting autonomy), with empirical work typically modeling SRL as the mediating bridge from need support/satisfaction to engagement and performance \cite{ryan2018self, jang2010engaging, sierens2009synergistic, miao2023teacher, bai2022effect}. Recent reviews further suggest that AI- and GenAI-supported learning activities increasingly operate through SRL processes, including goal setting, strategy selection, progress monitoring, information search, and feedback use, although the higher-education literature remains limited in explaining how AI shapes these SRL processes across phases of learning \cite{lan2025qualitative, qi2025systematic}.

As a driver of SRL, self-efficacy (SE) beliefs are also important in the study of AI impacts on the fulfillment of  psychological needs. SE represents beliefs about one’s capability to succeed in specific tasks and predicts not only SRL but effort, persistence, and strategy use \cite{bandura1997self}. Across secondary and higher-education settings, SE frequently mediates the association between need-supportive contexts/need satisfaction and engagement, deep learning, and achievement \cite{zimmerman2002becoming, bai2022effect, miao2023teacher, zhen2017mediating}. Within SDT, support for competence and autonomy fosters perceived effectance -- closely aligned with efficacy beliefs -- which energizes engagement and performance \cite{ryan2000self, ryan2018self}. Meta-analytic evidence also identifies academic SE as one of the strongest psychological correlates of university GPA. For these reasons and the fact that SE is fundamentally the confidence that drives the action inherent in SRL, this study includes self-efficacy as an additional covariate to SRL in the study of AI’s impact on psychological need satisfaction

\subsection{AI Support for Attention Challenges}
Students with attention-related difficulties face barriers that extend beyond knowledge acquisition, including difficulties sustaining focus, organizing work, and resisting distractions \cite{barkley1997behavioral, weyandt2006adhd, kofler2020working}. In response to these barriers, a range of AI interventions have been developed to address attentional and motivational barriers, particularly in K-12 contexts. Game-based approaches for children with attention deficit hyperactivity disorder (ADHD) often structure academic work into smaller, interactive units reinforced by feedback and rewards. \citet{souza2021use} examined a gamified learning environment that used task breakdown and point-based rewards to sustain attention and encourage steady progress, while \citet{ariyasena2024exploring} proposed a framework that integrates mini-games into the curriculum, using immediate feedback and achievement markers to maintain attention and motivation. Socially assistive robots have also been explored as supports for students with attention difficulties. Across studies of children and college students with ADHD, these systems function less as instructors and more as supportive companions or external cues that help redirect attention, structure tasks, and externalize regulation when focus drifts \cite{zuckerman2016kip3, lalwani2024productivity, o2024design, cervantes2023social}.
For example,robotic study partners such as KIP3 have been tested with students with ADHD as socially expressive feedback devices that provide immediate feedback for inattention or impulsivity events and help students regain focus \cite{zuckerman2016kip3}, and more recent work has explored similar study-companion roles for social robots that assist college students with ADHD in task prioritization, scheduling, and maintaining focus \cite{lalwani2025study}.
Beyond these robotic systems, other AI-supported modalities are also relevant to attention-related challenges. For example, systematic reviews suggest that VR-based interventions can support sustained attention and related outcomes in ADHD populations, particularly for children and adolescents \cite{goharinejad2022usefulness, romero2021effectiveness}. More recently, emerging HCI work has begun examining how adults with ADHD use ChatGPT in everyday life and how generative AI might support personalized learning interfaces for students with ADHD, although systems that leverage these insights are not yet established as ADHD-specific educational interventions \cite{pinto2025little, gunawardana2025enhancing}. More broadly, recent work shows that students with disabilities in higher education are already using generative AI tools such as chatbots, rewriting tools, and translation tools to work around barriers in academic writing, although concerns remain about accuracy, academic integrity, and unequal access due to subscription costs \cite{zhao2025use}.

Importantly, research grounded in SDT shows that inattention is associated with lower competence satisfaction and higher competence frustration, underscoring how unmet psychological needs can compound academic struggles in this population \cite{serrano2023adhd}. These challenges are particularly pronounced in higher education, where independent study and self-regulated learning are critical for success. Thus, while existing AI-based interventions highlight promising directions for supporting students with attention-related challenges, much of this work still does not directly measure psychological need fulfillment, often remains limited to small-scale or prototype deployments, and continues to emphasize K-12 rather than everyday higher education contexts, especially engineering education. As a result, much remains to be explored regarding how the type and level of psychological support AI provides differs for college students who face attentional challenges compared to those who do not. To address this gap in the research, this study also takes into consideration inattention in addition to everyday contexts and SRL and SE beliefs in exploring the relationship between AI tool use and psychological need support. 

\section{Methodology}
To answer our research questions, we designed and distributed a 10-15-minute online survey to engineering students on the Qualtrics platform in late May to early June 2025. The survey includes seven demographic questions as well as Likert-scaled items (eight constructs, 75 total items) intended to measure inattention, fulfillment of psychological needs (autonomy, competence, and relatedness), self-regulated learning, self-efficacy beliefs, and academic engagement.  Psychological needs fulfillment was measured both in a generalized context and in the context of specific AI tool use. The survey also included several multiple-choice and multi-select questions to identify factors that students perceive to impact their attention and motivation. And finally, open-ended questions were included in the survey to explore in greater depth how AI positively and negatively impacts students’ learning and focus. To address the research questions in this study, the analysis focused on demographic variables and the Likert-scale measures of inattention, psychological needs (autonomy, competence, and relatedness), self-regulated learning, and self-efficacy beliefs. A summary of survey measures relevant to this study is provided in Appendix \ref{sec:appendix}.  

\subsection{Study Recruitment and Respondents}
Undergraduate engineering students were recruited to complete the survey. The study was approved as exempt by the institutional review board (IRB) at a large public research university(IRB protocol number withheld for anonymous review) 
Recruitment was conducted during the last three weeks (Weeks 8 – 10) of the ten-week spring quarter in 2025 using two main strategies: 1) undergraduate advising teams distributed the survey invitation with mailing lists, and 2) professors shared the survey link with their classes, offering extra credit incentives for survey completion. To ensure that responses came from students at this large public research university, access to the survey required institutional single sign-on (SSO). To be eligible for the study, individuals had to provide consent and be at least 18 years old. Responses from participants who did not meet both criteria were excluded from the dataset but remained eligible to receive relevant incentives (extra credit or otherwise). The order of all Likert-scaled items was randomized to avoid fatigue effects.

A total of 335 students participated in the survey and completed the questionnaire in an average of 10-15 minutes. Six incomplete submissions were removed. Ninety-four submissions failed at least one instructional manipulation check (IMC) and were excluded (42 failed IMC 1; 52 failed IMC 2; see Appendix). IMC 1 was an attention check requiring selection of “Agree”; IMC 2 was an implausible statement for which endorsement indicated failure. Seven further submissions were excluded – one without recorded consent and six for low response quality (e.g., straight lining across Likert-type items). The remaining 228 participants were primarily electrical and computer engineering majors and represented multiple genders and races across all undergraduate academic standings (freshman, sophomore, junior, and senior; n = 227), as well as one participant enrolled in a bachelor’s and master’s combined degree program (n = 1). Chatbots (e.g., ChatGPT, Gemini, Claude, Microsoft Copilot, etc.) were the most common AI tool used by students (90.4\%) followed by adaptive learning platforms (14.0\%) and AI-powered tutoring (6.1\%). Among students who frequently used AI-based chatbots (n = 206), most were male (67.5\%), Asian (61.7\%), or juniors (41.3\%). Demographics for the survey subpopulation which used AI-based chatbots are detailed in Table \ref{tab:1}.

\begin{table}[!ht]
  \small
  \centering
    \caption{Demographics for Participants who used AI-based Chatbots (n = 206).}
  \resizebox{\columnwidth}{!}{
    \begin{tabular}{p{3cm} p{5cm} p{4cm} p{3cm}}
      \toprule
      \textbf{Demographic} & \textbf{Type} & \textbf{\# of Participants} & \textbf{Percentage (\%)}\\
      \midrule

      {\textbf{Gender}} 
        & Male & 139 & 67.48 \\
        & Female & 57 & 27.67 \\
        & Non-binary & 8 & 3.88 \\
        & Prefer not to say & 1 & 0.49 \\
        & Others & 1 & 0.49 \\
      \midrule

      {\textbf{Race}} 
        & White & 81 & 39.30 \\
        & Black or African American & 11 & 5.33 \\
        & American Indian or Alaska Native & 2 & 0.97 \\
        & Asian & 127 & 61.70 \\
        & Native Hawaiian or Pacific Islander & 4 & 1.94 \\
        & Others & 5 & 2.43 \\
      \midrule

      {\textbf{Academic Standings}} 
        & Freshman & 15 & 7.28 \\
        & Sophomore & 49 & 23.80 \\
        & Junior & 85 & 41.30 \\
        & Senior & 56 & 27.20 \\
        & Others & 1 & 0.49 \\
      \midrule

      {\textbf{Ethnicity}} 
        & Hispanic origin & 18 & 8.70 \\
        & Not Hispanic origin & 188 & 91.30 \\
      \midrule

      {\textbf{Academic Major}} 
        & Computer Science \& Engineering & 16 & 7.77 \\
        & Electrical \& Computer Engineering & 162 & 78.60 \\
        & Computer Engineering & 15 & 7.28 \\
        & Others & 13 & 6.31 \\
      \bottomrule
    \end{tabular}
  }
  \label{tab:1}
\end{table}

\subsection{Measures}
All items associated with this study used a 5-point Likert scale (1 = strongly disagree, 5 = strongly agree). Participants were instructed: “For the following questions, please respond with respect to your general experiences in classes and perceived improvements with the use of the specific AI tool in your major/discipline.” The survey included both satisfaction and frustration items for the psychological need constructs. Higher scores indicate greater satisfaction or frustration of autonomy, competence, or relatedness/belonging needs, higher self-regulated learning, or greater self-efficacy beliefs. Items associated with each of these scales were adapted from previously validated skills as follows:
\begin{itemize}
\item \textbf{Autonomy Needs} (adapted from \cite{chen2015basic, chen2015basic2, wilson2023engineering}): embodied perceptions of volition and choice in academic activities and included such items as “I feel a sense of choice and freedom in the curriculum I undertake.”.
\item \textbf{Competence Needs} (adapted from \cite{chen2015basic, chen2015basic2,  wilson2023engineering}): represented perceived capability to master and succeed at academic tasks and included such items as “I feel confident that I can successfully complete challenging tasks.”.
\item \textbf{Relatedness Needs} (adapted from \cite{chen2015basic, chen2015basic2, anderson2002factorial, wilson2023engineering}): referred to feelings of connection, care, and acceptance within one’s academic community and included such items as “I feel that the classmates and instructors I care about also care about me.”
\item \textbf{Self-Regulated Learning} (adapted from \cite{pintrich1991manual, wang2025artificial}): included planning, monitoring, help-seeking, and use of feedback to manage learning and included such items as “I feel comfortable connecting new knowledge with what I already know in my coursework.”
\item \textbf{Self-Efficacy Beliefs} (adapted from \cite{rigotti2008short, wilson2023engineering}): measured an individual’s belief in their capability to handle academic challenges and demands and included such items as “I can remain calm when facing difficulties because I trust my abilities.”
\end{itemize}
Inattention was measured using items from the Adult ADHD Self-Report Scale (ASRS) \cite{kessler2005world}. These items also used a 5-point Likert scale (1 = never, 5 = very often), with higher scores indicating lower level of attention.
\subsection{Analysis Procedure}
\begin{figure}[!ht]
  \centering
  \includegraphics[width=\columnwidth]{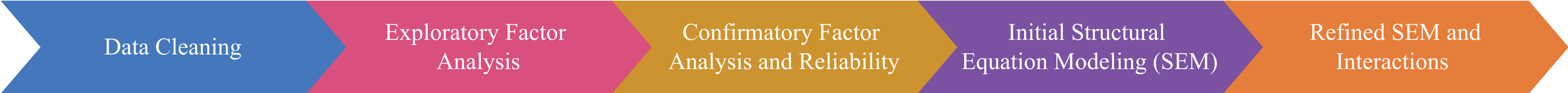}
  \caption{Overview of the analysis procedure. After response screening, exploratory factor analysis (EFA) was used to evaluate the dimensionality of the adapted measures, followed by confirmatory factor analysis (CFA) and reliability analysis. Structural equation modeling (SEM) was then used to estimate associations between baseline and perceived AI-related outcomes, after which the refined SEM and latent interaction models were estimated.}
  \Description{Overview of the analysis procedure.}
  \label{fig:overview_pipeline}
\end{figure}

Survey items related to the core measures in this study (competence, autonomy, relatedness, self-regulated learning, and self-efficacy) were first analyzed using confirmatory or exploratory factor analysis as appropriate to ensure construct validity and support scale reliability. The resulting scales were then integrated into structural equation modeling to explore the research questions that guided study. Figure \ref{fig:overview_pipeline} summarizes the overall analysis workflow.
\subsubsection{Factor Analysis}
Items on measuring self-efficacy beliefs, self-regulated learning, and fulfillment of psychological needs (competence, autonomy, and relatedness) were adapted from previous validated measures in higher education with minor wording changes. Items measuring inattention were identical to those used in previous studies (ADHD Self-Report scale for the inattention measures) \cite{kessler2005world}. Since these items were adapted and examined together in a new context, exploratory factor analysis (EFA) was conducted using the \texttt{psych} package \cite{revelle2026psych} in R (v 4.5.1) to evaluate their empirical dimensionality in the current sample \cite{knekta2019one}. For the competence, autonomy, and relatedness scales, reverse-worded items (n = 3) were removed prior to analysis so that retained indicators were aligned in the same direction and to reduce the likelihood that negative item wording would introduce wording-related method effects rather than substantively meaningful variance \cite{kam2023regular, dodeen2023effects}. All EFAs used maximum-likelihood estimation (MLE) with oblimin rotation \cite{jennrich1966rotation, carroll1953analytical}. An oblique rotation was selected because the latent constructs examined in this study (e.g., competence, autonomy, and relatedness) were theoretically expected to be correlated \cite{deci2000and, ryan2018self, fabrigar1999evaluating, costello2005best}. In contrast to commonly used orthogonal rotations such as varimax \cite{kaiser1958varimax}, the oblimin rotation permits nonzero correlations among factors and therefore provides a more appropriate representation of the underlying construct structure. Factorability was assessed using the Kaiser-Meyer-Olkin (KMO) measure and Bartlett’s test of sphericity \cite{kaiser1970second, bartlett1950tests}. The number of factors retained was determined by parallel analysis and interpretability. Items with EFA loadings greater than .30 were retained, consistent with a commonly used minimum threshold in exploratory analyses \cite{costello2005best}. However, final item retention was not based on loading magnitude alone. They were evaluated with respect to the overall factor pattern, interpretability, and conceptual coherence of the factor solution \cite{fabrigar1999evaluating, costello2005best}. Items were therefore retained only when they contributed to an interpretable construct and removed when their loadings were weak, isolated, or conceptually inconsistent with the factor on which they loaded.

Following the EFA analyses, confirmatory factor analysis (CFA) models were estimated in R (v. 4.5.1) using the \texttt{lavaan} package \cite{rosseel2012lavaan} to evaluate the measurement structures suggested by the EFA. Model fit was assessed using the Root Mean Square Error of Approximation (RMSEA), Comparative Fit Index (CFI), and Standardized Root Mean Square Residual (SRMR). RMSEA was selected to evaluate absolute model fit while accounting for model complexity, CFI to assess incremental fit relative to a null model, and SRMR to quantify the standardized discrepancy between observed and model-implied correlations. Together, these indices provide complementary perspectives on model fit while avoiding unnecessary redundancy across fit statistics \cite{hu1999cutoff}. Consistent with commonly used guidelines, CFI values of at least .90 were interpreted as indicating acceptable fit and values of at least .95 indicated strong fit \cite{hooper2008structural, marsh2004search}. For both RMSEA and SRMR, values below .05 indicated excellent fit, values between .05 and .08 indicated good fit, values between .08 and .10 indicated marginal fit, and values above .10 indicated poor fit \cite{hu1999cutoff}.

For constructs measured with only three indicators, standalone CFA models were not interpreted using global fit indices because such models are just-identified and therefore provide limited information regarding overall model fit \cite{wu2016identification}. Instead, these constructs were evaluated within the full structural equation modeling (SEM) framework, in which the overall model was over-identified and global fit statistics were informative. Following initial model evaluation, CFA models were re-estimated after removing items demonstrating weak empirical performance.

Internal consistency was assessed for each retained scale using Cronbach’s alpha and McDonald’s omega in the \texttt{psych} package \cite{revelle2026psych} in R (v. 4.5.1). Because several constructs in the final model were represented by only three indicators and equal factor loadings were not assumed, omega was reported alongside alpha to provide a reliability estimate. Reliability was calculated using the final retained items for each baseline and perceived construct.

\subsubsection{Effect Coding for Demographics}
Effect coding was used to represent categorical demographic variables for analysis in order to avoid implicit comparisons to majority populations (Table \ref{tab:effect-coded-demographics}). For race, the reference group coded as $-1$ was White, and Asian and underrepresented minorities (URM) were coded as $1$ in two separate demographic variables. The URM category included participants who identified as Black or African American, American Indian or Alaska Native, Native Hawaiian or Other Pacific Islander, due to the relatively small proportion of respondents in these groups. Gender was effect coded using a single female contrast, with Male as $-1$, Female as $1$, and nonbinary gender, other, and those who preferred not to answer coded as $0$. A separate gender contrast for nonbinary, other, and prefer-not-to-answer responses was not included because of the small number of respondents in these categories. Preliminary analyses indicated that academic standing had negligible predictive effect on the outcome variable so it was not included as a demographic category in the SEM models.  

\begin{table}[!ht]
  \small
  \centering
  \caption{Effect coded demographics variables.}
  \resizebox{\columnwidth}{!}{
    \begin{tabular}{p{3cm} p{3cm} p{3cm} p{3cm} p{3.2cm}}
      \toprule
      \textbf{Group} & \textbf{Asian Contrast} & \textbf{URM Contrast} & \textbf{Female Contrast} \\
      \midrule
      White, Male   & -1 & -1 & -1 \\
      White, Female & -1 & -1 &  1 \\
      White, Other  & -1 & -1 &  0 \\
      Asian, Male   &  1 &  0 & -1 \\
      Asian, Female &  1 &  0 &  1 \\
      Asian, Other  &  1 &  0 &  0 \\
      URM, Male     &  0 &  1 & -1 \\
      URM, Female   &  0 &  1 &  1 \\
      URM, Other    &  0 &  1 &  0 \\
      \bottomrule
    \end{tabular}
  }

  \vspace{2pt}
  \begin{minipage}{\columnwidth}
    \small
    * URM includes Black or African American, American Indian or Alaska Native, Native Hawaiian or Other Pacific Islander, and Other.\\
    ** Other gender includes Prefer not to answer,  Non-binary, and Other.
  \end{minipage}

  \label{tab:effect-coded-demographics}
\end{table}
\subsubsection{Structural Equation Modeling}
Structural equation modeling (SEM) was conducted in R with the \texttt{lavaan} package \cite{rosseel2012lavaan} to measure the associations between baseline latent constructs and perceived need fulfillment after AI use. The model specified baseline fulfillment of autonomy, competence, and relatedness needs, as well as baseline personal agency, as latent predictors of perceived autonomy, competence, and relatedness after AI use. Each perceived outcome was regressed on all baseline latent predictors while controlling for inattention and demographic covariates. Residual covariances were estimated among the perceived outcomes. The model was estimated using robust maximum likelihood. Following initial model estimation, standardized factor loadings and indicator-level explained variance were inspected to evaluate measurement quality. Items with weak standardized loadings and low explained variance were removed to improve measurement precision and model parsimony. The SEM was then re-estimated using the retained items. The resulting model was used as the primary model for interpreting structural associations.

Interaction effects were then estimated to test whether inattention moderated the association between baseline latent predictors, and perceived autonomy, perceived relief from competence frustration, and perceived relatedness. Before estimating interaction effects, the inattention score was mean-centered following standard practice in interaction analysis \cite{aguinis2017improving}. This means that zero represents average inattention, so the lower-order coefficients can be interpreted at the average level of inattention. The latent interaction effects were estimated in R with the modsem package with a product-indicator method \cite{lin2010structural, marsh2004structural}.
\section{Findings}
\subsection{Factor Analysis}
To evaluate the measurement structure of the adapted scales, EFA was first conducted on the combined items assessing self-regulated learning, self-efficacy beliefs, and fulfillment of competence, autonomy, and relatedness needs. For the competence, autonomy, and relatedness scales, reverse-worded items were removed before analysis to align the retained items in the same direction. For constructs with more than three items, the single-construct CFAs were conducted to assess the fit of the adapted measurement structure. These analyses were performed separately for measures in the traditional learning context and for perceived measures after AI use.
\subsubsection{EFA} For the baseline items, sampling adequacy was strong  ($KMO = .862$), and Bartlett’s test rejected the identity matrix ($\chi^2(171) = 1911.386$; $p < .001$) which indicates the correlation matrix is suitable for factor analysis. Parallel analysis suggested a four-factor structure. The resulting factor structure indicated that autonomy, competence, and relatedness each formed distinct factors (Table \ref{tab:efa_baseline}). In contrast, the self-regulated learning (SRL) and self-efficacy items loaded on a common factor, indicating empirical overlap between strategic learning behaviors and perceived self-beliefs about handling academic demands in this sample. This factor was labeled personal agency to reflect a combined sense of being able and ready to act on academic demands. Specifically, the retained items captured confidence in handling academic challenges, considering how to approach problems, developing one’s own ideas, and integrating new knowledge with prior understanding. This interpretation is consistent with prior work showing that self-efficacy and self-regulated learning are conceptually distinct but closely related: self-efficacy reflects learners’ beliefs about their capability to succeed, whereas self-regulated learning reflects the planning, monitoring, and strategic actions used to pursue academic goals \cite{bandura1997self, zimmerman2002becoming, pintrich2004conceptual}.

One of SRL items, “If I struggle with coursework, I know where to find help” (SR4), was removed from further analysis. Although SR4 showed a borderline loading on the relatedness factor ($\lambda = .359$), it did not load with the other SRL items and did not align conceptually with the factor on which it loaded. The retained items in the personal agency factor reflected internally oriented learning strategies and self-beliefs about handling academic demands, whereas SR4 emphasized awareness of external sources of help. This distinction is important because help-seeking is often treated as a specific form of self-regulated learning that relies on mobilizing external resources rather than on the internally managed planning, monitoring, and efficacy reflected in the retained items \cite{karabenick2011understanding, yang2023systematic}. SR4 also did not align conceptually with the relatedness factor, which reflected feelings of connection, acceptance, and care rather than resource awareness. SR4 was therefore removed on the basis of both borderline empirical fit and lack of conceptual coherence with the retained factors.
\begin{table}[!ht]
  \small
  \centering
  \caption{EFA for baseline psychological and motivational constructs.}
  \resizebox{\columnwidth}{!}{
    \begin{tabular}{p{3.5cm} p{1.5cm} p{2cm} p{2cm} p{2cm} p{2cm}}
      \toprule
      \textbf{Intended Constructs} & \textbf{Items} & \textbf{Factor 1} & \textbf{Factor 2} & \textbf{Factor 3} & \textbf{Factor 4}\\
      \midrule

      {\textbf{Autonomy}} 
        & AU1 &  &  &  & 0.416 \\
        & AU2 &  &  &  & 0.728 \\
        & AU3 &  &  &  & 0.776 \\
      \midrule

      {\textbf{Competence Frustration}} 
        & CO1 &  & 0.862 &  &  \\
        & CO2 &  & 0.420 &  &  \\
        & CO3 &  & 0.851 &  &  \\
        & CO4 &  & 0.771 &  &  \\
      \midrule

      {\textbf{Relatedness}} 
        & RE1 &  &  & 0.627 &  \\
        & RE2 &  &  & 0.725 &  \\
        & RE3 &  &  & 0.769 &  \\
      \midrule

      {\textbf{Self-regulated Learning}} 
        & SR1 & 0.840 &  &  &  \\
        & SR2 & 0.846 &  &  &  \\
        & SR3 & 0.760 &  &  &  \\
        & SR4 &  &  & 0.359 &  \\
        & SR5 & 0.623 &  &  &  \\
      \midrule

      {\textbf{Self-efficacy Beliefs}} 
        & SE1 & 0.748 &  &  &  \\
        & SE2 & 0.812 &  &  &  \\
        & SE3 & 0.842 &  &  &  \\
        & SE4 & 0.730 &  &  &  \\
      \bottomrule
    \end{tabular}
  }
  \label{tab:efa_baseline}
\end{table}

Similarly, for the perceived need-fulfillment items after AI use, sampling adequacy was strong  ($KMO = .823$), and Bartlett’s test rejected the identity matrix ($\chi^2(45) = 842.589$; $p < .001$). Parallel analysis suggested a three-factor structure. Factor loadings indicated that autonomy, competence, and relatedness each formed distinct factors (Table \ref{tab:efa_after_ai}). One competence item ACO2 did not meet the loading criterion on any factor ($\lambda = .107$, $.192$, and $.205$ across the three factors) and was therefore removed from further analysis.

The factor correlations produced by the oblimin-rotated EFA were inspected to evaluate the appropriateness of an oblique solution. Nonzero correlations among the retained factors would indicate that oblimin was more appropriate than an orthogonal rotation. For the baseline constructs, the factor correlation matrix showed generally small correlations among the retained factors, with coefficients ranging from $|r| = .022$ to $.321$ (Table \ref{tab:factor_correlation_baseline}). This pattern indicates that the baseline factors were related but empirically distinguishable, supporting the use of an oblique rotation. For the perceived psychological constructs after AI use, factor correlations were somewhat stronger, ranging from $r = .276$ to $r = .629$ (Table \ref{tab:factor_correlation_after_ai}). In particular, the correlation between Factors 1 and 3 was relatively high ($r = .629$), suggesting meaningful association among the perceived need-related outcomes while still supporting their separation as distinct factors.

\begin{table}[!ht]
  \small
  \centering
  \caption{EFA for psychological constructs after AI use.}
  \resizebox{\columnwidth}{!}{
    \begin{tabular}{p{4cm} p{2cm} p{2.5cm} p{2.5cm} p{2.5cm}}
      \toprule
      \textbf{Intended Constructs} & \textbf{Items} & \textbf{Factor 1} & \textbf{Factor 2} & \textbf{Factor 3}\\
      \midrule

      {\textbf{Autonomy}} 
        & AAU1 &  &  & 0.614 \\
        & AAU2 &  &  & 0.685 \\
        & AAU3 &  &  & 0.908 \\
      \midrule

      {\textbf{Competence Frustration}} 
        & ACO1 & 0.713 &  &  \\
        & ACO2 & & & \\
        & ACO3 & 0.782 &  &  \\
        & ACO4 & 0.818 &  &  \\
      \midrule

      {\textbf{Relatedness}} 
        & ARE1 &  & 0.875 &  \\
        & ARE2 &  & 0.719 &  \\
        & ARE3 &  & 0.769 &  \\
      \bottomrule
    \end{tabular}
  }
  \label{tab:efa_after_ai}
\end{table}

\begin{table}[!ht]
  \small
  \centering
  \caption{Factor correlation matrix for baseline constructs.}
  \resizebox{\columnwidth}{!}{
    \begin{tabular}{p{3cm} p{3cm} p{3cm} p{3cm} p{3cm}}
      \toprule
       & \textbf{Factor 1} & \textbf{Factor 2} & \textbf{Factor 3} & \textbf{Factor 4}\\
      \midrule

      {\textbf{Factor 1}} 
        & 1.000 & -0.321 & 0.094 & 0.278 \\
      \midrule

      {\textbf{Factor 2}} 
        & -0.321 & 1.000 & 0.022 & -0.049 \\
      \midrule

      {\textbf{Factor 3}} 
        & 0.094 & 0.022 & 1.000 & 0.295 \\
      \midrule

      {\textbf{Factor 4}} 
        & 0.278 & -0.049 & 0.295 & 1.000 \\
      \bottomrule
    \end{tabular}
  }
  \label{tab:factor_correlation_baseline}
\end{table}

\begin{table}[!ht]
  \small
  \centering
  \caption{Factor correlation matrix for psychological constructs after AI use.}
  \resizebox{\columnwidth}{!}{
    \begin{tabular}{p{3cm} p{3cm} p{3cm} p{3cm}}
      \toprule
       & \textbf{Factor 1} & \textbf{Factor 2} & \textbf{Factor 3}\\
      \midrule

      {\textbf{Factor 1}} 
        & 1.000 & 0.276 & 0.629 \\
      \midrule

      {\textbf{Factor 2}} 
        & 0.276 & 1.000 & 0.381 \\
      \midrule

      {\textbf{Factor 3}} 
        & 0.629 & 0.381 & 1.000 \\
      \bottomrule
    \end{tabular}
  }
  \label{tab:factor_correlation_after_ai}
\end{table}

The inattention items showed strong factorability ($KMO = .941$, $\chi^2(36) = 1063.232$; $p < .001$). Parallel analysis indicated a single-factor structure, and all nine items loaded strongly on that factor (Table \ref{tab:efa_inattention}).

\begin{table}[!ht]
  \small
  \centering
  \caption{EFA for Inattention.}
  \resizebox{\columnwidth}{!}{
    \begin{tabular}{p{3cm} p{1.5cm} p{1.5cm} p{1.5cm} p{1.5cm} p{1.5cm} p{1.5cm} p{1.5cm} p{1.5cm} p{1.5cm}}
      \toprule
      \textbf{Constructs} & \textbf{IA1} & \textbf{IA2} & \textbf{IA3} & \textbf{IA4} & \textbf{IA5} & \textbf{IA6} & \textbf{IA7} & \textbf{IA8} & \textbf{IA9}\\
      \midrule

      {\textbf{Inattention}} 
        & 0.755 & 0.735 & 0.780 & 0.792 & 0.824 & 0.746 & 0.721 & 0.690 & 0.737 \\
      \bottomrule
    \end{tabular}
  }
  \label{tab:efa_inattention}
\end{table}

\subsubsection{Single-construct CFAs} As suggested by the EFA results, single-construct CFAs were estimated for fulfillment of competence and personal agency and attention scale. All single-construct CFAs were specified using congeneric measurement models, allowing factor loadings to vary freely across indicators. This specification was chosen over imposing tau-equivalence constraints because it places no equality restrictions on indicator loadings. Although tau-equivalence models allow for estimation in low-indicator settings by constraining loadings to equality, such constraints impose a strong assumption of equal indicator contributions that was not supported by the observed loading patterns in Table \ref{tab:efa_baseline} and Table \ref{tab:efa_after_ai} \cite{graham2006congeneric}.

Table \ref{tab:cfa_fit_single_construct} below shows the results of CFA fit indices. In summary, the personal agency and attention scales showed good fit across indices. Fulfillment of competence also showed strong fit by CFI and SRMR, although RMSEA was slightly above the conventional cutoff. Previous work has shown that RMSEA can overstate misfit in models with small degrees of freedom, so this result was interpreted with caution and greater emphasis on CFI and SRMR results \cite{kenny2015performance, shi2022evaluating}. During subsequent SEM evaluation, one competence-frustration item was removed because of weak indicator performance. The single-construct CFA for competence was not re-estimated after this removal because the retained construct had three indicators and was subsequently evaluated within the full SEM. 

\begin{table}[!ht]
  \small
  \centering
  \caption{CFA fit indices for single-construct scales.}
  \resizebox{\columnwidth}{!}{
    \begin{tabular}{p{5cm} p{2cm} p{4cm} p{2.5cm} p{2.5cm}}
      \toprule
      \textbf{Constructs} & \textbf{df} & \textbf{RMSEA [90\% CI]} & \textbf{CFI} & \textbf{SRMR}\\
      \midrule

      \textbf{Competence Frustration} 
        & 2 & .088 [.000, .185] & 0.991 & 0.027 \\
      \midrule

      \textbf{Personal Agency} 
        & 20 & .077 [.047, .108] & 0.978 & 0.031 \\
      \midrule

      \textbf{Attention} 
        & 27 & .057 [.025, .085] & 0.983 & 0.031 \\
      \bottomrule
    \end{tabular}
  }
  \label{tab:cfa_fit_single_construct}
\end{table}

\subsection{Internal Consistency Reliability}
Table \ref{tab:internal_consistency} reports Cronbach’s alpha $\alpha$ and McDonald’s omega $\omega$ for the retained scales. Internal consistency was generally acceptable to strong across constructs. Baseline autonomy showed the lowest reliability ($\alpha = .692$, $\omega = .711$), but this was considered adequate for a brief three-item measure. Baseline competence and relatedness, as well as the perceived autonomy, competence, and relatedness scales, all showed good reliability, while personal agency showed the highest internal consistency ($\alpha = .925$, $\omega = .926$). The close correspondence between alpha and omega across scales further supports the reliability of the retained measures.
\begin{table}[!ht]
  \small
  \centering
  \caption{Internal Consistency of Retained Scales (Cronbach's $\alpha$ and McDonald's $\omega$).}
  \resizebox{\columnwidth}{!}{
    \begin{tabular}{p{4cm} p{3cm} p{4cm} p{4cm}}
      \toprule
      \textbf{Constructs} & \textbf{Number of Items} & \textbf{Cronbach's Alpha ($\alpha$)} & \textbf{McDonald's Omega ($\omega$)}\\
      \midrule

      {\textbf{Autonomy}} 
        & 3 & 0.692 & 0.711 \\
      \midrule

      {\textbf{Competence Frustration\textsuperscript{1}}} 
        & 3 & 0.871 & 0.873 \\
      \midrule

      {\textbf{Relatedness}} 
        & 3 & 0.760 & 0.763 \\
      \midrule

      {\textbf{Personal Agency}} 
        & 8 & 0.925 & 0.926 \\
      \midrule

      {\textbf{Perceived Autonomy}} 
        & 3 & 0.818 & 0.822 \\
      \midrule

      {\textbf{Perceived Competence}} 
        & 3 & 0.816 & 0.817 \\
      \midrule

      {\textbf{Perceived Relatedness}} 
        & 3 & 0.835 & 0.839 \\
      \midrule

      {\textbf{Attention}} 
        & 9 & 0.922 & 0.922 \\
      \bottomrule
    \end{tabular}
  }

  \vspace{0.5em}
  \begin{flushleft}
    \footnotesize{\textsuperscript{1} The internal consistency statistics are based on the model with CO2 removed.}
  \end{flushleft}
  \label{tab:internal_consistency}
\end{table}
\subsection{Descriptive Statistics}
Table \ref{tab:descriptive_statistics_ordinal} summarizes the descriptive statistics for the study variables. Among the baseline measures, students reported moderately high satisfaction of autonomy needs and relatedness, whereas competence frustration was closer to the scale midpoint. Personal agency was also relatively high. For the perceived outcomes resulting from AI Chatbot use, students reported moderate perceived relief from competence frustration and moderate perceived satisfaction of autonomy needs, whereas perceived satisfaction of relatedness needs was relatively lower. Across variables, skewness and kurtosis were modest, suggesting no substantial deviations from normality.

\begin{table}[!ht]
  \small
  \centering
  \caption{Descriptive Statistics (Ordinal Variables).}
  \resizebox{\columnwidth}{!}{
    \begin{tabular}{p{7cm} p{1.4cm} p{1.6cm} p{1.4cm} p{1.4cm} p{1.4cm} p{1.6cm} p{1.8cm}}
      \toprule
      \textbf{Measure} & \textbf{Mean} & \textbf{Median} & \textbf{Min} & \textbf{Max} & \textbf{SD} & \textbf{Skew} & \textbf{Kurtosis}\\
      \midrule

      \multicolumn{8}{c}{\textbf{Independent Variables}}\\
      \midrule

      Satisfaction of: 
        &  &  &  &  &  &  &  \\
      \hspace{1em} Autonomy Needs 
        & 3.45 & 3.33 & 1.33 & 5.00 & 0.76 & -0.22 & -0.26 \\
      \hspace{1em} Relatedness Needs 
        & 3.70 & 3.67 & 1.00 & 5.00 & 0.72 & -0.44 & 0.33 \\

      Frustration of: 
        &  &  &  &  &  &  &  \\
      \hspace{1em} Competence Needs\textsuperscript{1} 
        & 3.01 & 3.00 & 1.00 & 5.00 & 1.05 & -0.18 & -1.05 \\

      Inattention 
        & 17.48 & 17.00 & 0.00 & 36.00 & 9.25 & 0.14 & -0.27 \\

      Personal Agency 
        & 3.61 & 3.63 & 2.00 & 5.00 & 0.67 & 0.01 & -0.14 \\
      \midrule

      \multicolumn{8}{c}{\textbf{Dependent Variable}}\\
      \midrule

      Perceived Relief from Competence Frustration resulting from AI Chatbot Use 
        & 3.02 & 3.00 & 1.00 & 5.00 & 0.87 & -0.32 & -0.27 \\

      Perceived Satisfaction of Autonomy Needs resulting from AI Chatbot Use 
        & 2.96 & 3.00 & 1.00 & 5.00 & 0.84 & -0.02 & -0.09 \\

      Perceived Satisfaction of Relatedness Needs resulting from AI Chatbot Use 
        & 2.52 & 2.67 & 1.00 & 5.00 & 0.76 & -0.15 & -0.03 \\
      \bottomrule
    \end{tabular}
  }

  \vspace{0.5em}
  \begin{flushleft}
    \footnotesize{\textsuperscript{1} The descriptive statistics are based on the model with CO2 removed.}
  \end{flushleft}
  \label{tab:descriptive_statistics_ordinal}
\end{table}
\subsection{Structural Equation Modeling}
The structural equation model (SEM) fit the data well, with robust fit indices indicating strong overall fit. Examination of the measurement model identified one baseline competence item (CO2) with a low standardized factor loading ($.38$) and low explained variance ($.14$). The item was removed, and the model was re-estimated. The revised model showed improved fit (Table \ref{tab:model_fit_comparison}) while preserving the same overall pattern of structural relationships. All retained items loaded significantly on their intended latent constructs. Additionally, Akaike’s Information Criterion (AIC) also decreased which indicated improved relative model fit.

\begin{table}[!ht]
  \small
  \centering
  \caption{SEM Model Fit Comparison.}
  \resizebox{\columnwidth}{!}{
    \begin{tabular}{p{5cm} p{4cm} p{4cm}}
      \toprule
      \textbf{Fit Index} & \textbf{SEM 1} & \textbf{SEM 2 (CO2 removed)}\\
      \midrule

      $\chi^2$(df) 
        & 508.86 (399) & 457.07 (370) \\
      \midrule

      CFI 
        & .955 & .963 \\
      \midrule

      Tucker-Lewis Index (TLI) 
        & .948 & .958 \\
      \midrule

      RMSEA 
        & .037 & .034 \\
      \midrule

      SRMR 
        & .063 & .059 \\
      \midrule

      Akaike's Information Criterion (AIC) 
        & 12399.178 & 11829.470 \\
      \bottomrule
    \end{tabular}
  }
  \label{tab:model_fit_comparison}
\end{table}

\begin{figure}[!ht]
  \centering
  \includegraphics[width=\columnwidth]{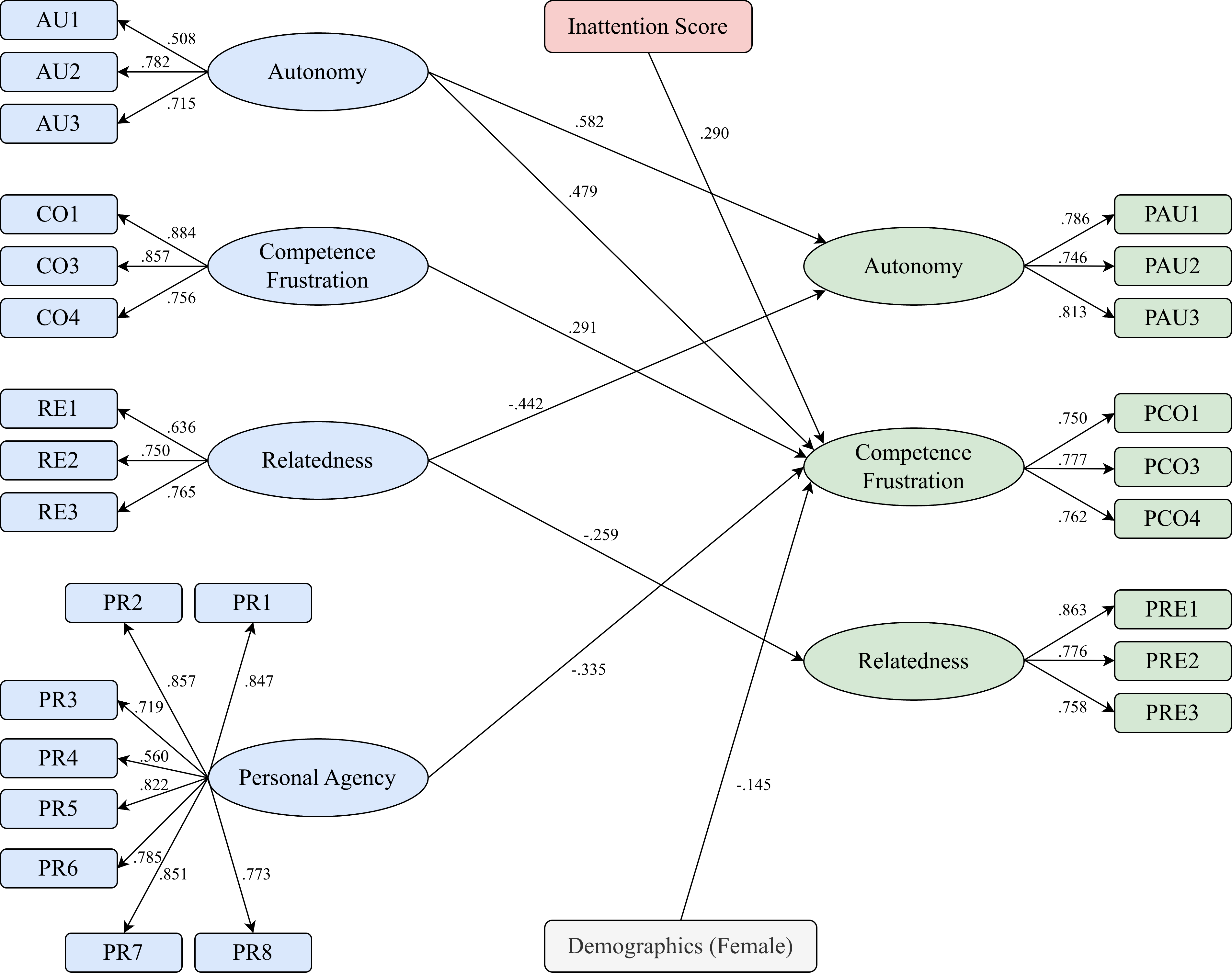}
  \caption{Structural equation model of baseline psychological constructs and perceived need fulfillment with AI use. Rectangles represent observed indicators and ovals represent latent constructs. Personal Agency corresponds to the combined baseline self-efficacy/self-regulated learning factor in the SEM. Standardized path coefficients are shown on the arrows. For readability, only statistically significant structural paths ($p < 0.05$) are included.}
  \Description{Structural equation model of baseline psychological constructs and perceived need fulfillment with AI use.}
  \label{fig:sem_model}
\end{figure}
Figure \ref{fig:sem_model} presents the final SEM with CO2 removed. Overall, the model explained 52.3\% of the variance in perceived competence, 37.5\% of the variance in perceived autonomy, and 9.7\% of the variance in perceived relatedness. Residual covariances among the three perceived outcomes remained significant. Students with higher baseline autonomy ($\beta = .479$, $p < .001$) and higher baseline competence frustration ($\beta = .291$, $p < .001$) were associated with greater perceived reduction in competence frustration following AI use. In contrast, higher baseline personal agency was associated with lower perceived relief from competence frustration when using AI ($\beta = -.335$, $p < .001$), and higher inattention was associated with greater perceived relief from competence frustration ($\beta = .290$, $p < .001$). Female-coded gender was also negatively associated with perceived competence related outcomes ($\beta = -.145$, $p = .037$). Higher baseline autonomy ($\beta = .582$, $p < .001$) was positively associated with perceived autonomy fulfillment from AI use, whereas higher baseline relatedness was negatively associated with perceived autonomy fulfillment from AI use ($\beta = -.442$, $p < .001$). Higher baseline relatedness was negatively associated with perceived relatedness satisfaction after AI chatbot use ($\beta = -.259$, $p = .014$), suggesting that students who initially reported stronger relatedness needs satisfaction perceived less additional relatedness support from AI chatbot use.
\subsubsection{Interaction effects} After controlling for demographics and all psychological baseline needs, two significant interaction effects emerged. Since inattention was mean-centered, the main effects of baseline psychological constructs should be interpreted as their associations with perceived outcomes at the average level of inattention. First, the interaction between baseline competence frustration and inattention significantly predicted perceived competence ($b = -.018$, $p = .003$). Students with greater baseline competence frustration tended to report greater relief from frustration, but this positive association weakened as inattention increased. In other words, higher inattention attenuated the association between baseline competence frustration and perceived relief from competence frustration. Second, the interaction between baseline autonomy and inattention was significant in predicting perceived autonomy ($b = -.033$, $p = .015$). This suggests that the positive association between baseline autonomy and perceived autonomy was also weaker for students with lower level of attention. No other tested interactions with inattention were statistically significant, including interactions involving baseline personal agency or relatedness.

\section{DISCUSSION}
\subsection{What is the impact of AI tool use on students’ psychological needs?}
Since this study relied on cross-sectional self-report data, this research question can only be answered in terms of students’ perceived AI-related benefits rather than causal change possible only with a true longitudinal study. Within this framing, students reported the greatest perceived relief from competence frustration following AI chatbot use, more moderate improvements in perceived satisfaction of autonomy needs, and the lowest perceived improvements in satisfaction of relatedness needs. This pattern is consistent with the SEM results, which explained substantially more variance in competence- and autonomy-related outcomes than in relatedness-related outcomes. Results are also consistent with SDT in that competence is the need most proximal to task performance -- supported by optimal challenge, clear structure, and effectance-relevant feedback -- so it tends to move on shorter horizons when learners receive immediate cues of progress \cite{ryan2018self, ryan2000self}. In contrast, autonomy and relatedness are more context-embedded; the instructional and social arrangements that strengthen them (e.g., teacher autonomy support combined with structure; peer involvement) typically accrue over weeks within a course, with longitudinal work showing that changes in engagement and autonomy-supportive teaching predict later-semester changes in need satisfaction and motivation \cite{jang2010engaging, jang2012longitudinal, reeve2014students}. Complementing SDT, mastery experiences are the primary driver of self-efficacy and update quickly when feedback is immediate. AI tool use provides such immediate feedback and enables increases in self-efficacy which, in turn, generates early perceived gains through competence-aligned signals \cite{pfitzner2016feel}. Taken together, these patterns suggest that AI chatbot use may be especially likely to influence competence needs in the short term, whereas autonomy- and relatedness-related benefits may depend more heavily on broader instructional and social context.

\subsection{How are these impacts affected by general satisfaction of psychological needs, self-regulated learning, and self-efficacy beliefs?}
The SEM results suggests that students’ baseline psychological needs and personal agency mattered more than demographic composition in shaping perceived psychological benefits from AI chatbot use. The only significant evidence of demographic influence in the SEM results occurred in relation to students’ gender. Female-coded students reported lower perceived relief from competence frustration. This pattern suggests lower perceived competence-related relief from AI chatbot use, rather than evidence of negative effects. This pattern may reflect broader gendered dynamics in engineering and computing contexts, where prior work has shown that female can report lower confidence or technology-related self-efficacy even when performance differences are minor. Although the observed gender effect was modest and should be interpreted cautiously, it highlights the need to examine whether AI-supported learning tools provide equally confidence-building and competence-supportive experiences across student groups \cite{beckwith2005effectiveness, beckwith2006tinkering, whitcomb2020mismatch}.

An examination of other potential influences on the perceived reduction in the frustration of competence needs via the use of AI chatbots revealed several important insights. While students’ baseline satisfaction of relatedness needs did not significantly predict perceived relief from competence frustration resulting from AI-tool use, baseline frustration of competence needs and baseline satisfaction of autonomy needs significantly and positively predicted these improvements. Self-determination theory (SDT) provides some insight into why these results are what they are. Competence is fostered by structure, optimal challenge, and informative feedback; thus when baseline competence frustration is high, scaffolding tends to have a larger impact on perceived relief from competence frustration \cite{ryan2020intrinsic, ryan2000self}. Instructional design theory similarly predicts diminishing returns and expertise reversal -- novices benefit more from guidance than advanced learners \cite{kirschner2010minimal, kalyuga2009expertise}. By structuring problem decomposition and offering just-in-time guidance, chatbots help learners convert high baseline competence frustration into measurable perceived relief from that frustration \cite{ma2025dbox, pardos2024chatgpt, phung2024automating, shabrina2023learning, vanzo2024gpt}.

In addition to baseline competence frustration, higher baseline autonomy also positively predicted perceived relief from competence frustration following AI chatbot use. Students who begin with a stronger sense of agency and choice are better positioned to use AI chatbots actively and strategically. This is consistent with SDT work showing that autonomy-supportive learning contexts are associated with stronger autonomous motivation, perceived competence, and related motivational outcomes \cite{niemiec2009autonomy, black2000effects}. Higher baseline autonomy may promote more agentic exploration and deeper processing, allowing students to leverage the chatbot’s open-ended affordances more fully. Research on agentic engagement similarly suggests that students learn more actively when they proactively contribute to and shape the learning process rather than passively receiving instruction \cite{reeve2014students, reeve2011agency}. Rather than passively receiving help, these students may be more willing to explore the solution space offered by the chatbot, redirect the interaction, and build on its responses in ways that make the tool more supportive of competence needs. By contrast, students with lower baseline autonomy may be less inclined to use the chatbot in these more exploratory ways, even when they have access to the same tool \cite{black2000effects, reeve2011agency}.

Moreover, baseline satisfaction of relatedness needs negatively predicted perceived relatedness satisfaction after AI chatbot use. This pattern suggests a relatedness-specific form of diminishing returns: students who already felt more socially connected in their learning environment may have had less unmet relatedness need for the chatbot to address. By contrast, students with lower baseline relatedness may have experienced AI chatbots as a more available, low-stakes, and responsive source of interaction, which could make the tool feel more relatedness-supportive. However, this result should not be interpreted as evidence that AI chatbots can replace peer, instructor, or classroom relationships. Rather, it suggests that perceived social support from AI may be most salient when students begin with weaker feelings of connection. Prior HCI work on relational agents and conversational agents shows that computational systems can evoke rapport, trust, and social presence, but also that such relational experiences depend strongly on design, user expectations, and the broader social context in which the system is used \cite{bickmore2005establishing, luger2016like, yang2021designing}. In educational settings, this finding points to a design opportunity: AI chatbots may be most valuable for relatedness when they help students feel accompanied, normalize help-seeking, or connect them back to human learning communities, rather than becoming a substitute for those communities \cite{wang2022co}.

In addition to the direct links between baseline psychological needs satisfaction and perceived relief from competence frustration resulting from AI Chatbot use, the variable labelled personal agency, which represents the combination of self-regulated learning and self-efficacy beliefs in this SEM model, significantly predicted perceived relief from competence frustration. Specifically, higher baseline personal agency was associated with lower perceived relief from competence frustration from AI Chatbot use. From an SDT perspective, structured, need-supportive guidance may be helpful especially when learners lack clear strategies whereas students already high in SRL or self-efficacy have less “headroom” for perceived relief from competence frustration \cite{ryan2020intrinsic, niemiec2009autonomy}. This “diminishing returns” pattern aligns with aptitude–treatment interaction and the expertise-reversal effect: guidance that powerfully helps novices yields smaller (or even negligible) marginal benefits for more expert or self-regulated learners \cite{kalyuga2007expertise, tetzlaff2025cornerstone}. It is also consistent with the “assistance dilemma,” which concerns how much guidance should be provided and suggests that heavier assistance is not always the most beneficial form of support for learners who are already better prepared \cite{mclaren2016efficiency}.

\subsection{How do students with attention-related challenges experience AI differently from other students?}
Higher inattention was associated with greater perceived relief from competence frustration following AI use. This pattern is consistent with the possibility that AI chatbots can provide forms of support that are especially helpful when students struggle to sustain focus, for example by breaking tasks into manageable steps and providing immediate feedback \cite{pergantis2025ai}. More broadly, attention-aware learning technologies and cognitive offloading research suggest that interactive systems can help learners manage executive-function demands by externalizing parts of the task and helping them re-engage when attention drifts \cite{risko2016cognitive, hutt2021breaking}. These affordances match with the cognitive load theory (which explains how limits of working memory shape learning) and create conditions in which attention-challenged learners can convert guidance into rapid competence frustration relief \cite{paas2003cognitive, sweller1988cognitive}. For instance, step-focused dialogue reduces extraneous cognitive load ("mind-wandering opportunities") and encourages cognitive offloading, narrowing the action space and keeping task goals salient \cite{risko2016cognitive, hutt2021breaking, smallwood2015science}. 

In addition to these baseline results regarding AI support for attention struggles, the interaction-effect models also suggest more nuanced patterns in that the benefits of AI tool use on inattention were not uniform. The positive association between baseline competence frustration and perceived relief from competence frustration weakened as inattention increased. Likewise, the positive association between baseline autonomy satisfaction and perceived autonomy satisfaction following AI use also weakened as inattention increased. In other words, higher inattention did not simply amplify AI-related benefits; rather, it changed how students’ starting motivational states translated into perceived outcomes. Taken together, these findings strongly suggest that a spectrum of inattention should be taken into account when designing AI tools for maximum psychological benefit.   

\subsection{Implications}
\subsubsection{AI Support Is Need-Specific Rather Than Uniform}
The findings strongly support that AI chatbot support should not be theorized or designed as a single undifferentiated motivational intervention \cite{ryan2020intrinsic, ballou2022self}. Instead, perceived AI-related benefits were strongest for relief from competence frustration, more moderate for autonomy, and weakest for relatedness. This pattern is consistent with SDT, which conceptualizes autonomy, competence, and relatedness as distinct psychological needs rather than as interchangeable indicators of general motivation \cite{knekta2019one}. More specifically, the findings suggest that AI chatbots may be especially effective at delivering competence-proximal support which emphasizes immediate structure, effectance-relevant feedback, and stepwise guidance that make progress easier to perceive in the moment. This interpretation aligns with SDT accounts of competence support and with learning-science work showing that highly structured guidance is often most beneficial when learners face uncertainty, lack clear strategies, or experience high initial difficulty \cite{ryan2020intrinsic, kirschner2010minimal, kalyuga2009expertise}. Meanwhile, the weaker relatedness findings suggest that conversational fluency alone should not be taken as evidence that AI supports belonging. In SDT terms, relatedness depends on experiences of mutual care, connection, and social embeddedness, which are typically produced through broader interpersonal and instructional arrangements rather than through task assistance alone \cite{ryan2020intrinsic, hew2023using, chiu2024teacher, kalyuga2009expertise}.

The results also suggest a more specific way to conceptualize learner heterogeneity in AI-supported learning. The strongest differences were associated less with demographic composition than with learners’ starting motivational profile, consisting of baseline competence frustration, autonomy, personal agency, and inattention.  These initial measures shaped how AI-related outcomes were experienced and underscores the need for theories of human-AI learning to pay greater attention to initial need states and capabilities as boundary conditions on perceived benefit, rather than assuming that the same AI affordances operate similarly for all students \cite{ryan2020intrinsic, wobbrock2018ability}. AI does not simply support learning, but more specifically, students’ need state, personal agency, and inattention influence how AI supports learning and can function as concrete adaptation targets in AI-mediated learning.

\subsubsection{Designing Needs-Aware, Attention-Aware AI Learning Tools}
\paragraph{\textit{Ability-Based, Needs-Aware Adaptation.}} AI learning tools should adapt to learners’ starting needs and abilities, rather than to demographic categories or a single fixed support policy. This is consistent with Ability-Based Design, which argues that interactive systems should be built around what users can do and how systems can adapt to those abilities, rather than framing users in terms of deficits \cite{wobbrock2011ability, wobbrock2018ability}. In the present case, the most relevant learner characteristics were not demographic categories but baseline competence frustration, autonomy, personal agency, and inattention. As reusable HCI design knowledge, this suggests that personalization in AI learning tools should be organized around learners’ current need state and attentional demand, not only around content difficulty or demographic grouping.

\paragraph{\textit{Contingent Tutoring Logic.}} Because higher baseline competence frustration was associated with greater perceived relief from competence frustration, whereas higher personal agency was associated with less such relief, AI chatbots should offer more scaffolding to learners with the most headroom and less redundant assistance to learners who are already well prepared. This follows prior work on the expertise-reversal effect, which shows that instructional guidance that benefits novices can become less useful, or even counterproductive, for more advanced learners \cite{kalyuga2009expertise, tetzlaff2025cornerstone}. A practical implication is to begin with lightweight diagnostics or early interaction cues and then adjust the amount of decomposition, prompting, worked guidance, and reflection support over the course of the interaction \cite{azevedo2022lessons, nye2014autotutor, graesser2016conversations, ma2025dbox}. Adaptation behavioral signals could include repeated help requests, long response latencies, repeated answer-seeking turns, and revision cycles without progress. These interaction signals need not function only as post hoc analytics; they can also serve as tailoring variables for dynamic support.

\paragraph{\textit{Attention-Aware Interface.}} Because inattention was associated with greater perceived relief from competence frustration, but also weakened the positive links from baseline competence frustration and autonomy to AI-related outcomes, the design goal should not be simply to provide “more help.” Instead, the interface should reduce executive-function demands by helping learners maintain task goals, externalize intermediate steps, and re-enter the task when attention drifts. This is consistent with cognitive offloading research, which shows that people use the environment to reduce internal cognitive demand, and with cognitive load theory, which emphasizes reducing unnecessary processing so that limited cognitive resources can be allocated to learning-relevant activity \cite{risko2016cognitive}. This yields a reusable HCI design implication: for attention-sensitive learning tasks, the key move is often to externalize control of task state rather than to increase the volume of generated explanation. Concretely, this points to design patterns such as visible next steps, saved state, checklists, bounded subtasks, progress indicators, and prompts that encourage learners to externalize partial reasoning before asking the model for the next step. These patterns are particularly relevant when the goal is not simply to find the right answer, but instead to sustain engagement with a multi-step task under attentional strain.

\paragraph{\textit{Relatedness-Supportive Guidance.}} Because relatedness showed the weakest benefits, AI learning tools should not assume that conversational polish is sufficient to support belonging. Instead, systems may need to route students toward human and peer support, for example, by helping them prepare questions for office hours, summarize where they are stuck before contacting an instructor, or identify an appropriate peer-collaboration next step. This is a generalizable HCI implication beyond the present dataset which advocates for an interface that  mediates access to real human beings when improvements in relatedness are a targeted outcome.  

\paragraph{\textit{Autonomy-Supportive Guidance.}} The autonomy findings suggest that learners benefit not only from structure, but from opportunities to remain agentic while using that structure. This implies that AI systems should make guidance negotiable rather than mandatory: learners should be able to ask for more or less detail, skip steps, request alternatives, or choose between worked guidance and open exploration. Such flexibility is consistent with autonomy-supportive environments, which support initiative and volition without abandoning structure \cite{reeve2013students, reeve2021autonomy}.

\subsubsection{Evaluating AI-Mediated Learning Experiences}
\paragraph{\textit{Measuring Mechanisms, Not Just Outcomes.}} The present findings also suggest that evaluation of AI learning tools should focus more directly on mechanisms of support, not only on broad end-state outcomes such as satisfaction or performance. In this study, the most interpretable effects were not generic “improvements,” but differential associations with perceived relief from competence frustration, autonomy-related outcomes, and relatedness-related outcomes. This implies that future HCI evaluations of AI-supported learning should measure specific psychological mechanisms, including need satisfaction, need frustration, personal agency, and attentional demands, rather than relying solely on post-use attitudes. The distinction between need satisfaction and need frustration is theoretically meaningful in SDT and has been explicitly operationalized in prior measurement work \cite{chen2015basic, chen2015basic2}.

\paragraph{\textit{Evaluating for Whom AI Support Works.}} The results suggest that average support claims may overlook important differences across learners, so it is important to explore who benefits from AI most. The current pattern suggests that the benefits of AI support depend on learners’ starting state, including competence frustration, personal agency, autonomy, and inattention. Accordingly, evaluations should routinely test moderation and heterogeneous effects, especially when the proposed value of the system lies in adaptive or personalized support \cite{kalyuga2009expertise, wobbrock2018ability}.

\paragraph{\textit{Moving Beyond Cross-Sectional Inference.}} Because the present analyses are cross-sectional and self-reported, they are best interpreted as evidence about perceived AI-related benefits rather than causal change. This limitation is not unique to the present study. In fact, methodological work has shown that cross-sectional mediation analyses can generate biased estimates of processes that unfold over time. In terms of AI-supported learning, this implies that stronger evidence will come from longitudinal, pre/post, or classroom-embedded experimental designs that can test both outcome change and the mechanisms through which interface features shape that change. Such evaluations should also incorporate more objective learning measures to distinguish perceived relief or support from actual learning gains, for example, correctness on comparable problems, performance on knowledge master and transfer, and course-based outcomes such as quiz or assignment scores \cite{maxwell2007bias, maxwell2011bias}.

\paragraph{\textit{In-Process Measures for Adaptation.}} The system evaluation should include process telemetry alongside outcome measures. If the proposed value of AI support lies in reducing extraneous load, externalizing executive control, or helping learners sustain progress, then these claims should be tested with interaction-level evidence such as hint requests, time between steps, revision cycles, use of saved state, and adherence to suggested task decomposition. Such measures would help bridge the gap between interface design and psychological outcomes \cite{risko2016cognitive}. They can also serve to support dynamic adaptation, allowing the system to detect struggle, disengagement, or overload and then adjust the level or form of scaffolding in response. This points to a more novel HCI research agenda where testing whether moment-to-moment adaptive scaffolding improves both need-related experiences and objective learning outcomes relative to static prompting strategies.

\subsection{Limitations}
Our findings should be interpreted in light of several limitations. First, the sample was drawn from a large public research university in the United States and was concentrated in engineering contexts, which limits generalizability to other institutions and other disciplines. Although the sample is disproportionately male, it is broadly consistent with engineering demographics both locally and nationally \cite{asee2024profiles}. Second, the study relied heavily on self-report measures; despite IMCs and quality screens, such measures are vulnerable to response and misinterpretation.  In particular, the inattention measure reflects perceived attentional difficulties rather than a clinical diagnosis, which may have introduced significant variability in true inattention. Third, participants differed in AI experience and in the specific tools, versions, and tasks they used even within the “chatbot” category, so unobserved heterogeneity in use frequency and intensity and context may remain. Fourth, several constructs were measured using adapted items, and some scales were ultimately represented by only three retained indicators. Reverse-worded need items were removed to reduce wording-related method effects, and one competence-frustration item was removed during SEM refinement because of weak indicator performance. These decisions improved measurement coherence, but they also narrowed construct coverage and may have reduced comparability with the original validated instruments. Finally, the data were observational and primarily cross-sectional. The SEM and interaction effects should therefore be interpreted as associational patterns in perceived AI-related benefits rather than as evidence of causal change. In addition, the study did not include objective or longer-term learning outcomes, such as correctness on comparable problems, transfer, retention, or course performance, limiting claims about whether perceived psychological benefits translate into sustained academic gains.

\section{Conclusion}
This study shows that the perceived psychological benefits of AI chatbot use are need-specific rather than uniform. In this sample of engineering students, AI chatbot use was associated most strongly with perceived relief from competence frustration, more moderately with autonomy-related benefits, and only weakly with relatedness-related benefits. Baseline competence frustration, autonomy, personal agency, and inattention mattered more than demographic composition in explaining these outcomes, and inattention altered how baseline competence frustration and autonomy translated into perceived AI-related benefits. These findings suggest that AI learning tools should be designed as needs- and attention- aware systems that diagnose learners’ starting state, provide contingent scaffolding, externalize task states when attention drifts, and preserve learner agency through negotiable guidance. Future evaluations should also move beyond broad endpoint measures and more directly assess need-related mechanisms, heterogeneous effects, and in-process signals that can support dynamic adaptation. However, since this study is limited by a single site, self-report scales, an observational design, and short-term outcomes, which constrain generality and causal claims, further research may be necessary to solidify these claims into actionable design guidelines.

To address these limitations and test the generality of the present findings, future work will:
\begin{itemize}
\item Go beyond chatbots and examine whether the need-specific pattern observed here generalizes across other AI learning tools reported in the survey, including adaptive learning platforms, AI-powered tutoring, and immersive applications, whose affordances may differentially support competence, autonomy, and relatedness.
\item Use longitudinal and pre/post designs that track both perceived need-related outcomes and more objective learning outcomes, such as correctness on comparable problems, transfer, retention, and course performance, in order to test whether short-term psychological benefits translate into durable learning gains.
\item Co-design and evaluate a discipline-specific, dynamically adaptive system for ECE that uses in-process telemetry, such as help requests, response latency, and adherence to suggested decomposition, to tailor scaffolding to learners’ need state, personal agency, and attentional demands, and then examine how these design principles transfer to other STEM disciplines and institutional contexts.
\end{itemize}

\bibliographystyle{ACM-Reference-Format}
\bibliography{ADE-refs}

@techreport{asee2024profiles,
  author = {{American Society for Engineering Education}},
  title = {Profiles of Engineering and Engineering Technology, 2023},
  institution = {American Society for Engineering Education},
  address = {Washington, DC},
  year = {2024},
  url  = {https://ira.asee.org/by-the-numbers/}
}

@article{tyack2024self,
  title={Self-determination theory and HCI games research: Unfulfilled promises and unquestioned paradigms},
  author={Tyack, April and Mekler, Elisa D},
  journal={ACM Transactions on Computer-Human Interaction},
  volume={31},
  number={3},
  pages={1--74},
  year={2024},
  publisher={ACM New York, NY}
}

@article{maxwell2011bias,
  title={Bias in cross-sectional analyses of longitudinal mediation: Partial and complete mediation under an autoregressive model},
  author={Maxwell, Scott E and Cole, David A and Mitchell, Melissa A},
  journal={Multivariate behavioral research},
  volume={46},
  number={5},
  pages={816--841},
  year={2011},
  publisher={Taylor \& Francis}
}

@article{reeve2021autonomy,
  title={Autonomy-supportive teaching: Its malleability, benefits, and potential to improve educational practice},
  author={Reeve, Johnmarshall and Cheon, Sung Hyeon},
  journal={Educational psychologist},
  volume={56},
  number={1},
  pages={54--77},
  year={2021},
  publisher={Taylor \& Francis}
}

@article{reeve2013students,
  title={How students create motivationally supportive learning environments for themselves: The concept of agentic engagement.},
  author={Reeve, Johnmarshall},
  journal={Journal of educational psychology},
  volume={105},
  number={3},
  pages={579},
  year={2013},
  publisher={American Psychological Association}
}

@article{wobbrock2018ability,
  title={Ability-based design},
  author={Wobbrock, Jacob O and Gajos, Krzysztof Z and Kane, Shaun K and Vanderheiden, Gregg C},
  journal={Communications of the ACM},
  volume={61},
  number={6},
  pages={62--71},
  year={2018},
  publisher={ACM New York, NY, USA}
}

@article{pergantis2025ai,
  title={AI chatbots and cognitive control: enhancing executive functions through chatbot interactions: a systematic review},
  author={Pergantis, Pantelis and Bamicha, Victoria and Skianis, Charalampos and Drigas, Athanasios},
  journal={Brain Sciences},
  volume={15},
  number={1},
  pages={47},
  year={2025},
  publisher={MDPI}
}

@article{mclaren2016efficiency,
  title={The efficiency of worked examples compared to erroneous examples, tutored problem solving, and problem solving in computer-based learning environments},
  author={McLaren, Bruce M and Van Gog, Tamara and Ganoe, Craig and Karabinos, Michael and Yaron, David},
  journal={Computers in Human Behavior},
  volume={55},
  pages={87--99},
  year={2016},
  publisher={Elsevier}
}

@article{kalyuga2007expertise,
  title={Expertise reversal effect and its implications for learner-tailored instruction},
  author={Kalyuga, Slava},
  journal={Educational psychology review},
  volume={19},
  number={4},
  pages={509--539},
  year={2007},
  publisher={Springer}
}

@inproceedings{wang2022co,
  title={Co-designing AI agents to support social connectedness among online learners: functionalities, social characteristics, and ethical challenges},
  author={Wang, Qiaosi and Jing, Shan and Goel, Ashok K},
  booktitle={Proceedings of the 2022 ACM Designing Interactive Systems Conference},
  pages={541--556},
  year={2022}
}

@article{bickmore2005establishing,
  title={Establishing and maintaining long-term human-computer relationships},
  author={Bickmore, Timothy W and Picard, Rosalind W},
  journal={ACM Transactions on Computer-Human Interaction (TOCHI)},
  volume={12},
  number={2},
  pages={293--327},
  year={2005},
  publisher={ACM New York, NY, USA}
}

@inproceedings{luger2016like,
  title={" Like Having a Really Bad PA" The Gulf between User Expectation and Experience of Conversational Agents},
  author={Luger, Ewa and Sellen, Abigail},
  booktitle={Proceedings of the 2016 CHI conference on human factors in computing systems},
  pages={5286--5297},
  year={2016}
}

@article{whitcomb2020mismatch,
  title={A mismatch between self-efficacy and performance: Undergraduate women in engineering tend to have lower self-efficacy despite earning higher grades than men},
  author={Whitcomb, Kyle M and Kalender, Z Yasemin and Nokes-Malach, Timothy J and Schunn, Christian D and Singh, Chandralekha},
  journal={arXiv preprint arXiv:2003.06006},
  year={2020}
}

@inproceedings{beckwith2006tinkering,
  author    = {Beckwith, L. and Kissinger, C. and Burnett, M. and Wiedenbeck, S. and Lawrance, J. and Blackwell, A. and Cook, C.},
  title     = {Tinkering and Gender in End-User Programmers' Debugging},
  booktitle = {Proceedings of the SIGCHI Conference on Human Factors in Computing Systems},
  series    = {CHI '06},
  year      = {2006},
  pages     = {231--240},
  publisher = {ACM},
  address   = {New York, NY, USA}
}

@inproceedings{beckwith2005effectiveness,
  author    = {Beckwith, L. and Burnett, M. and Wiedenbeck, S. and Cook, C. and Sorte, S. and Hastings, M.},
  title     = {Effectiveness of End-User Debugging Software Features: Are There Gender Issues?},
  booktitle = {Proceedings of the SIGCHI Conference on Human Factors in Computing Systems},
  series    = {CHI '05},
  year      = {2005},
  pages     = {869--878},
  publisher = {ACM},
  address   = {New York, NY, USA}
}

@article{shi2022evaluating,
  title={Evaluating SEM model fit with small degrees of freedom},
  author={Shi, Dexin and DiStefano, Christine and Maydeu-Olivares, Alberto and Lee, Taehun},
  journal={Multivariate behavioral research},
  volume={57},
  number={2-3},
  pages={179--207},
  year={2022},
  publisher={Taylor \& Francis}
}

@article{kenny2015performance,
  title={The performance of RMSEA in models with small degrees of freedom},
  author={Kenny, David A and Kaniskan, Burcu and McCoach, D Betsy},
  journal={Sociological methods \& research},
  volume={44},
  number={3},
  pages={486--507},
  year={2015},
  publisher={Sage Publications Sage CA: Los Angeles, CA}
}

@article{graham2006congeneric,
  title={Congeneric and (essentially) tau-equivalent estimates of score reliability: What they are and how to use them},
  author={Graham, James M},
  journal={Educational and psychological measurement},
  volume={66},
  number={6},
  pages={930--944},
  year={2006},
  publisher={Sage Publications Sage CA: Thousand Oaks, CA}
}

@article{yang2023systematic,
  title={A Systematic Review of Studies Exploring Help-Seeking Strategies in Online Learning Environments.},
  author={Yang, Fan and Stefaniak, Jill},
  journal={Online Learning},
  volume={27},
  number={1},
  pages={107--126},
  year={2023},
  publisher={ERIC}
}

@article{anderson2002factorial,
  title={Factorial and criterion validity of scores of a measure of belonging in youth development programs},
  author={Anderson-Butcher, Dawn and Conroy, David E},
  journal={Educational and psychological measurement},
  volume={62},
  number={5},
  pages={857--876},
  year={2002},
  publisher={Sage Publications Sage CA: Thousand Oaks, CA}
}

@article{rigotti2008short,
  title={A short version of the occupational self-efficacy scale: Structural and construct validity across five countries},
  author={Rigotti, Thomas and Schyns, Birgit and Mohr, Gisela},
  journal={Journal of Career Assessment},
  volume={16},
  number={2},
  pages={238--255},
  year={2008},
  publisher={Sage Publications Sage CA: Los Angeles, CA}
}

@Manual{revelle2026psych,
  title = {psych: Procedures for Psychological, Psychometric, and Personality Research},
  author = {{William Revelle}},
  organization = {Northwestern University},
  address = {Evanston, Illinois},
  year = {2026},
  note = {R package version 2.6.5},
  url = {https://CRAN.R-project.org/package=psych}
}

@article{rosseel2012lavaan,
  title={lavaan: An R package for structural equation modeling},
  author={Rosseel, Yves},
  journal={Journal of statistical software},
  volume={48},
  pages={1--36},
  year={2012}
}

@article{marsh2004structural,
  title={Structural equation models of latent interactions: evaluation of alternative estimation strategies and indicator construction.},
  author={Marsh, Herbert W and Wen, Zhonglin and Hau, Kit-Tai},
  journal={Psychological methods},
  volume={9},
  number={3},
  pages={275},
  year={2004},
  publisher={American Psychological Association}
}

@article{lin2010structural,
  title={Structural equation models of latent interactions: Clarification of orthogonalizing and double-mean-centering strategies},
  author={Lin, Guan-Chyun and Wen, Zhonglin and Marsh, Herbert W and Lin, Huey-Shyan},
  journal={Structural Equation Modeling},
  volume={17},
  number={3},
  pages={374--391},
  year={2010},
  publisher={Taylor \& Francis}
}

@article{aguinis2017improving,
  title={Improving our understanding of moderation and mediation in strategic management research},
  author={Aguinis, Herman and Edwards, Jeffrey R and Bradley, Kyle J},
  journal={Organizational research methods},
  volume={20},
  number={4},
  pages={665--685},
  year={2017},
  publisher={Sage Publications Sage CA: Los Angeles, CA}
}

@article{marsh2004search,
  title={In search of golden rules: Comment on hypothesis-testing approaches to setting cutoff values for fit indexes and dangers in overgeneralizing Hu and Bentler's (1999) findings},
  author={Marsh, Herbert W and Hau, Kit-Tai and Wen, Zhonglin},
  journal={Structural equation modeling},
  volume={11},
  number={3},
  pages={320--341},
  year={2004},
  publisher={Taylor \& Francis}
}

@misc{hooper2008structural,
  title={Structural Equation Modelling: Guidelines for Determining Model Fit Structural equation modelling: guidelines for determining model fit. Dublin Institute of Technology ARROW@ DIT, 6 (1), 53--60},
  author={Hooper, D and Coughlan, J and Mullen, MR and Mullen, J and Hooper, D and Coughlan, J and Mullen, MR},
  year={2008}
}

@article{wu2016identification,
  title={Identification of confirmatory factor analysis models of different levels of invariance for ordered categorical outcomes},
  author={Wu, Hao and Estabrook, Ryne},
  journal={Psychometrika},
  volume={81},
  number={4},
  pages={1014--1045},
  year={2016},
  publisher={Cambridge University Press \& Assessment}
}

@article{fabrigar1999evaluating,
  title={Evaluating the use of exploratory factor analysis in psychological research.},
  author={Fabrigar, Leandre R and Wegener, Duane T and MacCallum, Robert C and Strahan, Erin J},
  journal={Psychological methods},
  volume={4},
  number={3},
  pages={272},
  year={1999},
  publisher={American Psychological Association}
}

@article{costello2005best,
  title={Best practices in exploratory factor analysis: Four recommendations for getting the most from your analysis},
  author={Costello, Anna B and Osborne, Jason},
  journal={Practical assessment, research, and evaluation},
  volume={10},
  number={1},
  year={2005},
  publisher={University of Massachusetts Amherst Libraries}
}

@article{bartlett1950tests,
  title={Tests of significance in factor analysis.},
  author={Bartlett, Maurice S},
  journal={British journal of psychology},
  year={1950},
  publisher={British Psychological Society}
}

@article{kaiser1970second,
  title={A second generation little jiffy},
  author={Kaiser, Henry F},
  journal={Psychometrika},
  volume={35},
  number={4},
  pages={401--415},
  year={1970},
  publisher={Springer-Verlag}
}

@article{kaiser1958varimax,
  title={The varimax criterion for analytic rotation in factor analysis},
  author={Kaiser, Henry F},
  journal={Psychometrika},
  volume={23},
  number={3},
  pages={187--200},
  year={1958},
  publisher={Springer-Verlag}
}

@article{carroll1953analytical,
  title={An analytical solution for approximating simple structure in factor analysis},
  author={Carroll, John B},
  journal={Psychometrika},
  volume={18},
  number={1},
  pages={23--38},
  year={1953},
  publisher={Springer}
}

@article{jennrich1966rotation,
  title={Rotation for simple loadings},
  author={Jennrich, Robert I and Sampson, PF},
  journal={Psychometrika},
  volume={31},
  number={3},
  pages={313--323},
  year={1966},
  publisher={Springer-Verlag}
}

@article{dodeen2023effects,
  title={The effects of changing negatively worded items to positively worded items on the reliability and the factor structure of psychological scales},
  author={Dodeen, Hamzeh},
  journal={Journal of Psychoeducational Assessment},
  volume={41},
  number={3},
  pages={298--310},
  year={2023},
  publisher={SAGE Publications Sage CA: Los Angeles, CA}
}

@article{kam2023regular,
  title={Why do regular and reversed items load on separate factors? Response difficulty vs. item extremity},
  author={Kam, Chester Chun Seng},
  journal={Educational and Psychological Measurement},
  volume={83},
  number={6},
  pages={1085--1112},
  year={2023},
  publisher={Sage Publications Sage CA: Los Angeles, CA}
}

@article{knekta2019one,
  title={One size doesn’t fit all: Using factor analysis to gather validity evidence when using surveys in your research},
  author={Knekta, Eva and Runyon, Christopher and Eddy, Sarah},
  journal={CBE—Life Sciences Education},
  volume={18},
  number={1},
  pages={rm1},
  year={2019},
  publisher={American Society for Cell Biology}
}

@inproceedings{gunawardana2025enhancing,
  title={Enhancing Adaptive Personalized Learning Interfaces with Generative AI for Individuals with ADHD},
  author={Gunawardana, Rashmi and Perera, Michelle and Lakshika, Rashini and Ranasinghe, Chaminda and Karunanayaka, Kasun},
  booktitle={Proceedings of the 16th International Conference of Human-Computer Interaction (HCI) Design \& Research},
  pages={131--143},
  year={2025}
}

@inproceedings{pinto2025little,
  title={" A little bit of a life raft”--Exploring the Use and Experiences of ChatGPT as a Support Tool among Adults with ADHD},
  author={Pinto, Anika and Quilter, Emily and Vasikaran, Jerusaa and Koerner, Sarah and Chimonas, Tefkros and Nielsen, Emily Esther and Marshall, Paul and O'Hara, Kenton},
  booktitle={Proceedings of the 37th Australian Conference on Human-Computer Interaction},
  pages={50--67},
  year={2025}
}

@article{romero2021effectiveness,
  title={Effectiveness of virtual reality-based interventions for children and adolescents with ADHD: A systematic review and meta-analysis},
  author={Romero-Ayuso, Dulce and Toledano-Gonz{\'a}lez, Abel and Rodr{\'\i}guez-Mart{\'\i}nez, Mar{\'\i}a Del Carmen and Arroyo-Castillo, Palma and Trivi{\~n}o-Ju{\'a}rez, Jos{\'e} Mat{\'\i}as and Gonz{\'a}lez, Pascual and Ariza-Vega, Patrocinio and Del Pino Gonzalez, Antonio and Segura-Fragoso, Antonio},
  journal={Children},
  volume={8},
  number={2},
  pages={70},
  year={2021},
  publisher={Mdpi}
}

@article{goharinejad2022usefulness,
  title={The usefulness of virtual, augmented, and mixed reality technologies in the diagnosis and treatment of attention deficit hyperactivity disorder in children: an overview of relevant studies},
  author={Goharinejad, Saeideh and Goharinejad, Samira and Hajesmaeel-Gohari, Sadrieh and Bahaadinbeigy, Kambiz},
  journal={BMC psychiatry},
  volume={22},
  number={1},
  pages={4},
  year={2022},
  publisher={Springer}
}

@article{cervantes2023social,
  title={Social robots and brain--computer interface video games for dealing with attention deficit hyperactivity disorder: A systematic review},
  author={Cervantes, Jos{\'e}-Antonio and L{\'o}pez, Sonia and Cervantes, Salvador and Hern{\'a}ndez, Aribei and Duarte, Heiler},
  journal={Brain Sciences},
  volume={13},
  number={8},
  pages={1172},
  year={2023},
  publisher={MDPI}
}

@inproceedings{lalwani2024productivity,
  title={Productivity coachbot: a social robot coach for university students with adhd},
  author={Lalwani, Himanshi and Elgarf, Maha and Salam, Hanan},
  booktitle={A3DE, ACM/IEEE International Conference on Human-Robot Interaction. IEEE},
  year={2024}
}

@article{zhao2025use,
  title={The use of generative AI by students with disabilities in higher education},
  author={Zhao, Xin and Cox, Andrew and Chen, Xuanning},
  journal={The Internet and Higher Education},
  volume={66},
  pages={101014},
  year={2025},
  publisher={Elsevier}
}

@article{qi2025systematic,
  title={A systematic literature review on designing self-regulated learning using generative artificial intelligence and its future research directions},
  author={Qi, XIA and Liu, Qian and Tlili, Ahmed and Thomas, KF},
  journal={Computers \& education},
  pages={105465},
  year={2025},
  publisher={Elsevier}
}

@article{lan2025qualitative,
  title={A qualitative systematic review on AI empowered self-regulated learning in higher education},
  author={Lan, Min and Zhou, Xiaofeng},
  journal={npj Science of Learning},
  volume={10},
  number={1},
  pages={21},
  year={2025},
  publisher={Nature Publishing Group UK London}
}

@inproceedings{yang2021designing,
  title={Designing conversational agents: A self-determination theory approach},
  author={Yang, Xi and Aurisicchio, Marco},
  booktitle={Proceedings of the 2021 CHI Conference on Human Factors in Computing Systems},
  pages={1--16},
  year={2021}
}

@inproceedings{ballou2022self,
  title={Self-determination theory in HCI: shaping a research agenda},
  author={Ballou, Nick and Deterding, Sebastian and Tyack, April and Mekler, Elisa D and Calvo, Rafael A and Peters, Dorian and Villalobos-Z{\'u}{\~n}iga, Gabriela and Turkay, Selen},
  booktitle={CHI conference on human factors in computing systems extended abstracts},
  pages={1--6},
  year={2022}
}

@article{chen2015basic,
  title={Basic psychological need satisfaction, need frustration, and need strength across four cultures},
  author={Chen, Beiwen and Vansteenkiste, Maarten and Beyers, Wim and Boone, Liesbet and Deci, Edward L and Van der Kaap-Deeder, Jolene and Duriez, Bart and Lens, Willy and Matos, Lennia and Mouratidis, Athanasios and others},
  journal={Motivation and emotion},
  volume={39},
  number={2},
  pages={216--236},
  year={2015},
  publisher={Springer}
}

@article{chen2015basic2,
  title={Basic psychological need satisfaction and frustration scale},
  author={Chen, Beiwen and Vansteenkiste, Maarten and Beyers, Wim and Boone, Liesbet and Deci, Edward L and Van der Kaap-Deeder, Jolene and Duriez, Bart and Lens, Willy and Matos, Lennia and Mouratidis, Athanasios and others},
  journal={Motivation and Emotion},
  year={2015}
}

@article{pintrich1991manual,
  title={A manual for the use of the Motivated Strategies for Learning Questionnaire (MSLQ).},
  author={Pintrich, Paul R and others},
  year={1991},
  publisher={ERIC}
}

@article{maxwell2007bias,
  title={Bias in cross-sectional analyses of longitudinal mediation.},
  author={Maxwell, Scott E and Cole, David A},
  journal={Psychological methods},
  volume={12},
  number={1},
  pages={23},
  year={2007},
  publisher={American Psychological Association}
}

@article{wobbrock2011ability,
  title={Ability-based design: Concept, principles and examples},
  author={Wobbrock, Jacob O and Kane, Shaun K and Gajos, Krzysztof Z and Harada, Susumu and Froehlich, Jon},
  journal={ACM Transactions on Accessible Computing (TACCESS)},
  volume={3},
  number={3},
  pages={1--27},
  year={2011},
  publisher={ACM New York, NY, USA}
}

@article{sweller1988cognitive,
  title={Cognitive load during problem solving: Effects on learning},
  author={Sweller, John},
  journal={Cognitive science},
  volume={12},
  number={2},
  pages={257--285},
  year={1988},
  publisher={Elsevier}
}

@article{paas2003cognitive,
  title={Cognitive load theory and instructional design: Recent developments},
  author={Paas, Fred and Renkl, Alexander and Sweller, John},
  journal={Educational psychologist},
  volume={38},
  number={1},
  pages={1--4},
  year={2003},
  publisher={Taylor \& Francis}
}

@inproceedings{hutt2021breaking,
  title={Breaking out of the lab: Mitigating mind wandering with gaze-based attention-aware technology in classrooms},
  author={Hutt, Stephen and Krasich, Kristina and R. Brockmole, James and K. D'Mello, Sidney},
  booktitle={Proceedings of the 2021 CHI conference on human factors in computing systems},
  pages={1--14},
  year={2021}
}

@article{smallwood2015science,
  title={The science of mind wandering: Empirically navigating the stream of consciousness},
  author={Smallwood, Jonathan and Schooler, Jonathan W},
  journal={Annual review of psychology},
  volume={66},
  number={1},
  pages={487--518},
  year={2015},
  publisher={Annual Reviews}
}

@article{risko2016cognitive,
  title={Cognitive offloading},
  author={Risko, Evan F and Gilbert, Sam J},
  journal={Trends in cognitive sciences},
  volume={20},
  number={9},
  pages={676--688},
  year={2016},
  publisher={Elsevier}
}

@article{tetzlaff2025cornerstone,
  title={A cornerstone of adaptivity--A meta-analysis of the expertise reversal effect},
  author={Tetzlaff, Leonard and Simonsmeier, Bianca and Peters, Tabea and Brod, Garvin},
  journal={Learning and Instruction},
  volume={98},
  pages={102142},
  year={2025},
  publisher={Elsevier}
}

@article{shabrina2023learning,
  title={Learning Problem Decomposition-Recomposition with Data-Driven Chunky Parsons Problems within an Intelligent Logic Tutor.},
  author={Shabrina, Preya and Mostafavi, Behrooz and Tithi, Sutapa Dey and Chi, Min and Barnes, Tiffany},
  journal={International Educational Data Mining Society},
  year={2023},
  publisher={ERIC}
}

@article{pardos2024chatgpt,
  title={ChatGPT-generated help produces learning gains equivalent to human tutor-authored help on mathematics skills},
  author={Pardos, Zachary A and Bhandari, Shreya},
  journal={Plos one},
  volume={19},
  number={5},
  pages={e0304013},
  year={2024},
  publisher={Public Library of Science San Francisco, CA USA}
}

@article{vanzo2024gpt,
  title={GPT-4 as a homework tutor can improve student engagement and learning outcomes},
  author={Vanzo, Alessandro and Chowdhury, Sankalan Pal and Sachan, Mrinmaya},
  journal={arXiv preprint arXiv:2409.15981},
  year={2024}
}

@inproceedings{phung2024automating,
  title={Automating human tutor-style programming feedback: Leveraging gpt-4 tutor model for hint generation and gpt-3.5 student model for hint validation},
  author={Phung, Tung and P{\u{a}}durean, Victor-Alexandru and Singh, Anjali and Brooks, Christopher and Cambronero, Jos{\'e} and Gulwani, Sumit and Singla, Adish and Soares, Gustavo},
  booktitle={Proceedings of the 14th learning analytics and knowledge conference},
  pages={12--23},
  year={2024}
}

@inproceedings{ma2025dbox,
  title={Dbox: Scaffolding algorithmic programming learning through learner-llm co-decomposition},
  author={Ma, Shuai and Wang, Junling and Zhang, Yuanhao and Ma, Xiaojuan and Wang, April Yi},
  booktitle={Proceedings of the 2025 CHI Conference on Human Factors in Computing Systems},
  pages={1--20},
  year={2025}
}

@article{reeve2011agency,
  title={Agency as a fourth aspect of students’ engagement during learning activities},
  author={Reeve, Johnmarshall and Tseng, Ching-Mei},
  journal={Contemporary educational psychology},
  volume={36},
  number={4},
  pages={257--267},
  year={2011},
  publisher={Elsevier}
}

@article{black2000effects,
  title={The effects of instructors' autonomy support and students' autonomous motivation on learning organic chemistry: A self-determination theory perspective},
  author={Black, Aaron E and Deci, Edward L},
  journal={Science education},
  volume={84},
  number={6},
  pages={740--756},
  year={2000},
  publisher={Wiley Online Library}
}

@article{kirschner2010minimal,
  title={Why minimal guidance during instruction does not work: An analysis of the failure of constructivist},
  author={Kirschner, Paul A and Sweller, John and Clark, Richard E and Kirschner, PA and Clark, RE},
  journal={Based Teaching Work: An Analysis of the Failure of Constructivist, Discovery, Problem-Based, Experiential, and Inquiry-Based Teaching,(November 2014)},
  pages={37--41},
  year={2010}
}

@incollection{kalyuga2009expertise,
  title={The expertise reversal effect},
  author={Kalyuga, Slava},
  booktitle={Managing cognitive load in adaptive multimedia learning},
  pages={58--80},
  year={2009},
  publisher={IGI Global Scientific Publishing}
}

@article{pfitzner2016feel,
  title={Why do I feel more confident? Bandura's sources predict preservice teachers' latent changes in teacher self-efficacy},
  author={Pfitzner-Eden, Franziska},
  journal={Frontiers in psychology},
  volume={7},
  pages={1486},
  year={2016},
  publisher={Frontiers Media SA}
}

@article{reeve2014students,
  title={Students’ classroom engagement produces longitudinal changes in classroom motivation.},
  author={Reeve, Johnmarshall and Lee, Woogul},
  journal={Journal of educational psychology},
  volume={106},
  number={2},
  pages={527},
  year={2014},
  publisher={American Psychological Association}
}

@article{jang2012longitudinal,
  title={Longitudinal test of self-determination theory's motivation mediation model in a naturally occurring classroom context.},
  author={Jang, Hyungshim and Kim, Eun Joo and Reeve, Johnmarshall},
  journal={Journal of Educational psychology},
  volume={104},
  number={4},
  pages={1175},
  year={2012},
  publisher={American Psychological Association}
}

@article{kessler2005world,
  title={The World Health Organization Adult ADHD Self-Report Scale (ASRS): a short screening scale for use in the general population},
  author={Kessler, Ronald C and Adler, Lenard and Ames, Minnie and Demler, Olga and Faraone, Steve and Hiripi, EVA and Howes, Mary J and Jin, Robert and Secnik, Kristina and Spencer, Thomas and others},
  journal={Psychological medicine},
  volume={35},
  number={2},
  pages={245--256},
  year={2005},
  publisher={Cambridge University Press}
}

@article{hu1999cutoff,
  title={Cutoff criteria for fit indexes in covariance structure analysis: Conventional criteria versus new alternatives},
  author={Hu, Li-tze and Bentler, Peter M},
  journal={Structural equation modeling: a multidisciplinary journal},
  volume={6},
  number={1},
  pages={1--55},
  year={1999},
  publisher={Taylor \& Francis}
}

@inproceedings{wilson2023engineering,
  title={Engineering CAReS: Measuring Basic Psychological Needs in the Engineering Workplace},
  author={Wilson, Denise and VanAntwerp, Jennifer J and Misra, Shruti},
  booktitle={2023 ASEE Annual Conference \& Exposition},
  year={2023}
}

@article{wang2025artificial,
  title={Artificial intelligence in higher education: The impact of need satisfaction on artificial intelligence literacy mediated by self-regulated learning strategies},
  author={Wang, Kai and Cui, Wencheng and Yuan, Xue},
  journal={Behavioral Sciences},
  volume={15},
  number={2},
  pages={165},
  year={2025},
  publisher={MDPI}
}

@article{zhen2017mediating,
  title={The mediating roles of academic self-efficacy and academic emotions in the relation between basic psychological needs satisfaction and learning engagement among Chinese adolescent students},
  author={Zhen, Rui and Liu, Ru-De and Ding, Yi and Wang, Jia and Liu, Ying and Xu, Le},
  journal={Learning and individual differences},
  volume={54},
  pages={210--216},
  year={2017},
  publisher={Elsevier}
}

@article{sierens2009synergistic,
  title={The synergistic relationship of perceived autonomy support and structure in the prediction of self-regulated learning},
  author={Sierens, Eline and Vansteenkiste, Maarten and Goossens, Luc and Soenens, Bart and Dochy, Filip},
  journal={British journal of educational psychology},
  volume={79},
  number={1},
  pages={57--68},
  year={2009},
  publisher={Wiley Online Library}
}

@article{miao2023teacher,
  title={Teacher autonomy support influence on online learning engagement: the mediating roles of self-efficacy and self-regulated learning},
  author={Miao, Jia and Ma, Li},
  journal={Sage Open},
  volume={13},
  number={4},
  pages={21582440231217737},
  year={2023},
  publisher={SAGE Publications Sage CA: Los Angeles, CA}
}

@article{bai2022effect,
  title={Effect of teacher autonomy support on the online self-regulated learning of students during COVID-19 in China: The chain mediating effect of parental autonomy support and students’ self-efficacy},
  author={Bai, Xuemei and Gu, Xiaoqing},
  journal={Journal of Computer Assisted Learning},
  volume={38},
  number={4},
  pages={1173--1184},
  year={2022},
  publisher={Wiley Online Library}
}

@article{crede2011meta,
  title={A meta-analytic review of the Motivated Strategies for Learning Questionnaire},
  author={Cred{\'e}, Marcus and Phillips, L Alison},
  journal={Learning and individual differences},
  volume={21},
  number={4},
  pages={337--346},
  year={2011},
  publisher={Elsevier}
}

@article{richardson2012psychological,
  title={Psychological correlates of university students' academic performance: a systematic review and meta-analysis.},
  author={Richardson, Michelle and Abraham, Charles and Bond, Rod},
  journal={Psychological bulletin},
  volume={138},
  number={2},
  pages={353},
  year={2012},
  publisher={American Psychological Association}
}

@article{chen2022effectiveness,
  title={The effectiveness of self-regulated learning (SRL) interventions on L2 learning achievement, strategy employment and self-efficacy: A meta-analytic study},
  author={Chen, Jing},
  journal={Frontiers in Psychology},
  volume={13},
  pages={1021101},
  year={2022},
  publisher={Frontiers Media SA}
}

@article{theobald2021self,
  title={Self-regulated learning training programs enhance university students’ academic performance, self-regulated learning strategies, and motivation: A meta-analysis},
  author={Theobald, Maria},
  journal={Contemporary Educational Psychology},
  volume={66},
  pages={101976},
  year={2021},
  publisher={Elsevier}
}

@article{dignath2008can,
  title={How can primary school students learn self-regulated learning strategies most effectively?: A meta-analysis on self-regulation training programmes},
  author={Dignath, Charlotte and Buettner, Gerhard and Langfeldt, Hans-Peter},
  journal={Educational Research Review},
  volume={3},
  number={2},
  pages={101--129},
  year={2008},
  publisher={Elsevier}
}

@article{nye2014autotutor,
  title={AutoTutor and family: A review of 17 years of natural language tutoring},
  author={Nye, Benjamin D and Graesser, Arthur C and Hu, Xiangen},
  journal={International Journal of Artificial Intelligence in Education},
  volume={24},
  number={4},
  pages={427--469},
  year={2014},
  publisher={Springer}
}

@article{graesser2016conversations,
  title={Conversations with AutoTutor help students learn},
  author={Graesser, Arthur C},
  journal={International Journal of Artificial Intelligence in Education},
  volume={26},
  number={1},
  pages={124--132},
  year={2016},
  publisher={Springer}
}

@article{aleven2016help,
  title={Help helps, but only so much: Research on help seeking with intelligent tutoring systems},
  author={Aleven, Vincent and Roll, Ido and McLaren, Bruce M and Koedinger, Kenneth R},
  journal={International Journal of Artificial Intelligence in Education},
  volume={26},
  number={1},
  pages={205--223},
  year={2016},
  publisher={Springer}
}

@article{pintrich2004conceptual,
  title={A conceptual framework for assessing motivation and self-regulated learning in college students},
  author={Pintrich, Paul R},
  journal={Educational psychology review},
  volume={16},
  number={4},
  pages={385--407},
  year={2004},
  publisher={Springer}
}

@article{luo2025design,
  title={Design and assessment of AI-based learning tools in higher education: a systematic review},
  author={Luo, Jihao and Zheng, Chenxu and Yin, Jiamin and Teo, Hock Hai},
  journal={International Journal of Educational Technology in Higher Education},
  volume={22},
  number={1},
  pages={42},
  year={2025},
  publisher={Springer}
}

@inproceedings{duan2022designing,
  title={Designing a learning analytics dashboard to provide students with actionable feedback and evaluating its impacts},
  author={Duan, Xiaojing and Wang, Chaoli and Rouamba, Guieswende},
  booktitle={Proceedings of International Conference on Computer Supported Education},
  year={2022}
}

@inproceedings{jin2024teach,
  title={Teach ai how to code: Using large language models as teachable agents for programming education},
  author={Jin, Hyoungwook and Lee, Seonghee and Shin, Hyungyu and Kim, Juho},
  booktitle={Proceedings of the 2024 CHI Conference on Human Factors in Computing Systems},
  pages={1--28},
  year={2024}
}

@inproceedings{basileo2024role,
  title={The role of self-efficacy, motivation, and perceived support of students' basic psychological needs in academic achievement},
  author={Basileo, Lindsey D and Otto, Barbara and Lyons, Merewyn and Vannini, Natalie and Toth, Michael D},
  booktitle={Frontiers in education},
  volume={9},
  pages={1385442},
  year={2024},
  organization={Frontiers Media SA}
}

@article{schunk2020motivation,
  title={Motivation and social cognitive theory},
  author={Schunk, Dale H and DiBenedetto, Maria K},
  journal={Contemporary educational psychology},
  volume={60},
  pages={101832},
  year={2020},
  publisher={Elsevier}
}

@incollection{schunk2016self,
  title={Self-efficacy theory in education},
  author={Schunk, Dale H and DiBenedetto, Maria K},
  booktitle={Handbook of motivation at school},
  pages={34--54},
  year={2016},
  publisher={Routledge}
}

@article{ryan2020intrinsic,
  title={Intrinsic and extrinsic motivation from a self-determination theory perspective: Definitions, theory, practices, and future directions},
  author={Ryan, Richard M and Deci, Edward L},
  journal={Contemporary educational psychology},
  volume={61},
  pages={101860},
  year={2020},
  publisher={Elsevier}
}

@inproceedings{atcheson2025d,
  title={" I'd Never Actually Realized How Big An Impact It Had Until Now": Perspectives of University Students with Disabilities on Generative Artificial Intelligence},
  author={Atcheson, Alex and Khan, Omar and Siemann, Brian and Jain, Anika and Karahalios, Karrie},
  booktitle={Proceedings of the 2025 CHI Conference on Human Factors in Computing Systems},
  pages={1--22},
  year={2025}
}

@inproceedings{glazko2023autoethnographic,
  title={An autoethnographic case study of generative artificial intelligence's utility for accessibility},
  author={Glazko, Kate S and Yamagami, Momona and Desai, Aashaka and Mack, Kelly Avery and Potluri, Venkatesh and Xu, Xuhai and Mankoff, Jennifer},
  booktitle={Proceedings of the 25th International ACM SIGACCESS Conference on Computers and Accessibility},
  pages={1--8},
  year={2023}
}

@inproceedings{tcherdakoff2025designing,
  title={Designing for Neurodiversity in Academia: Addressing Challenges and Opportunities in Human-Computer Interaction},
  author={Tcherdakoff, Nathalie Alexandra Penglin and Stangroome, Grace Jane and Milton, Ashlee and Holloway, Catherine and Cecchinato, Marta E and Nonnis, Antonella and Eagle, Tessa and Al Thani, Dena and Hong, Hwajung and Williams, Rua Mae},
  booktitle={Proceedings of the Extended Abstracts of the CHI Conference on Human Factors in Computing Systems},
  pages={1--5},
  year={2025}
}

@article{alvarez2023systematic,
  title={A systematic review of actions aimed at university students with ADHD},
  author={{\'A}lvarez-Godos, Mar{\'\i}a and Ferreira, Camino and Vieira, Mar{\'\i}a-Jos{\'e}},
  journal={Frontiers in psychology},
  volume={14},
  pages={1216692},
  year={2023},
  publisher={Frontiers Media SA}
}

@article{sedgwick2022university,
  title={University students with attention deficit hyperactivity disorder (ADHD): a consensus statement from the UK Adult ADHD Network (UKAAN)},
  author={Sedgwick-M{\"u}ller, Jane A and M{\"u}ller-Sedgwick, Ulrich and Adamou, Marios and Catani, Marco and Champ, Rebecca and Gudj{\'o}nsson, G{\'\i}sli and Hank, Dietmar and Pitts, Mark and Young, Susan and Asherson, Philip},
  journal={BMC psychiatry},
  volume={22},
  number={1},
  pages={292},
  year={2022},
  publisher={Springer}
}

@article{scheres2021adhd,
  title={Do ADHD symptoms, executive function, and study strategies predict temporal reward discounting in college students with varying levels of ADHD symptoms? A pilot study},
  author={Scheres, Anouk and Solanto, Mary V},
  journal={Brain sciences},
  volume={11},
  number={2},
  pages={181},
  year={2021},
  publisher={MDPI}
}

@article{staley2024attention,
  title={Attention-deficit/hyperactivity disorder diagnosis, Treatment, and telehealth use in adults—National center for health statistics rapid surveys system, United States, October--November 2023},
  author={Staley, Brooke S},
  journal={MMWR. Morbidity and Mortality Weekly Report},
  volume={73},
  year={2024}
}

@article{zimmerman2002becoming,
  title={Becoming a self-regulated learner: An overview},
  author={Zimmerman, Barry J},
  journal={Theory into practice},
  volume={41},
  number={2},
  pages={64--70},
  year={2002},
  publisher={Taylor \& Francis}
}

@book{bandura1997self,
  title={Self-efficacy: The exercise of control},
  author={Bandura, Albert},
  volume={11},
  year={1997},
  publisher={Freeman}
}

@article{karabenick2011understanding,
  title={Understanding and facilitating self-regulated help seeking},
  author={Karabenick, Stuart A and Dembo, Myron H},
  journal={New directions for teaching and learning},
  volume={2011},
  number={126},
  pages={33--43},
  year={2011},
  publisher={Wiley Online Library}
}

@book{ryan2018self,
  title={Self-Determination Theory: Basic Psychological Needs in Motivation, Development, and Wellness},
  author={Ryan, R.M. and Deci, E.L.},
  isbn={9781462538966},
  lccn={2016046662},
  series={Self-determination Theory: Basic Psychological Needs in Motivation, Development, and Wellness},
  url={https://books.google.com/books?id=th5rDwAAQBAJ},
  year={2018},
  publisher={Guilford Publications}
}

@article{serrano2023adhd,
  title={ADHD and psychological need fulfillment in college students},
  author={Serrano, Judah W and Abu-Ramadan, Tamara M and Vasko, John M and Leopold, Daniel R and Canu, Will H and Willcutt, Erik G and Hartung, Cynthia M},
  journal={Journal of Attention Disorders},
  volume={27},
  number={8},
  pages={912--924},
  year={2023},
  publisher={SAGE Publications Sage CA: Los Angeles, CA}
}

@inproceedings{zuckerman2016kip3,
author = {Zuckerman, Oren and Hoffman, Guy and Kopelman-Rubin, Daphne and Klomek, Anat Brunstein and Shitrit, Noa and Amsalem, Yahav and Shlomi, Yaron},
title = {KIP3: Robotic Companion as an External Cue to Students with ADHD},
year = {2016},
isbn = {9781450335829},
publisher = {Association for Computing Machinery},
address = {New York, NY, USA},
url = {https://doi.org/10.1145/2839462.2856535},
doi = {10.1145/2839462.2856535},
booktitle = {Proceedings of the TEI '16: Tenth International Conference on Tangible, Embedded, and Embodied Interaction},
pages = {621–626},
numpages = {6},
location = {Eindhoven, Netherlands},
series = {TEI '16}
}

@inproceedings{lalwani2025study,
  title={A study companion for productivity: exploring the role of a social robot for college students with ADHD},
  author={Lalwani, Himanshi and Saleh, Mira and Salam, Hanan},
  booktitle={2025 20th ACM/IEEE International Conference on Human-Robot Interaction (HRI)},
  pages={1438--1442},
  year={2025},
  organization={IEEE}
}

@inproceedings{o2024design,
  title={Design and evaluation of a socially assistive robot schoolwork companion for college students with adhd},
  author={O'Connell, Amy and Banga, Ashveen and Ayissi, Jennifer and Yaminrafie, Nikki and Ko, Ellen and Le, Andrew and Cislowski, Bailey and Mataric, Maja},
  booktitle={Proceedings of the 2024 ACM/IEEE International Conference on Human-Robot Interaction},
  pages={533--541},
  year={2024}
}

@inproceedings{ariyasena2024exploring,
  title={Exploring the Utility of Gamified Learning to Support Academic Progress among ADHD-Diagnosed Children},
  author={Ariyasena, Prathibha Sathyanjalee and Munasinghe, Nimthara Nuwangi Bandara and Chandra, Christan Fryer Cooraydas and Dissanayake, Prashanthi Anushika and Samarakoon, Uthpala and Tissera, Wishalya},
  booktitle={Proceedings of the 2024 Sixteenth International Conference on Contemporary Computing},
  pages={170--176},
  year={2024}
}

@inproceedings{souza2021use,
  title={The use of games as an educational aid to students with ADHD: a framework proposal},
  author={Souza, Tain{\'a} and Silva, M{\^o}nica and Alves, Renato and Moura, Waldir},
  booktitle={2021 International Conference on Advanced Learning Technologies (ICALT)},
  pages={333--335},
  year={2021},
  organization={IEEE}
}

@article{kofler2020working,
  title={Working memory and short-term memory deficits in ADHD: A bifactor modeling approach.},
  author={Kofler, Michael J and Singh, Leah J and Soto, Elia F and Chan, Elizabeth SM and Miller, Caroline E and Harmon, Sherelle L and Spiegel, Jamie A},
  journal={Neuropsychology},
  volume={34},
  number={6},
  pages={686},
  year={2020},
  publisher={American Psychological Association}
}

@article{weyandt2006adhd,
  title={ADHD in college students},
  author={Weyandt, Lisa L and DuPaul, George},
  journal={Journal of attention disorders},
  volume={10},
  number={1},
  pages={9--19},
  year={2006},
  publisher={Sage Publications Sage CA: Thousand Oaks, CA}
}

@article{barkley1997behavioral,
  title={Behavioral inhibition, sustained attention, and executive functions: constructing a unifying theory of ADHD.},
  author={Barkley, Russell A},
  journal={Psychological bulletin},
  volume={121},
  number={1},
  pages={65},
  year={1997},
  publisher={American Psychological Association}
}

@inproceedings{blattgerste2022motivational,
  title={Motivational benefits and usability of a handheld Augmented Reality game for anatomy learning},
  author={Blattgerste, Jonas and Franssen, Jannik and Arztmann, Michaela and Pfeiffer, Thies},
  booktitle={2022 IEEE International Conference on Artificial Intelligence and Virtual Reality (AIVR)},
  pages={266--274},
  year={2022},
  organization={IEEE}
}

@article{elford2022fostering,
  title={Fostering motivation toward chemistry through augmented reality educational escape activities. A self-determination theory approach},
  author={Elford, Daniel and Lancaster, Simon J and Jones, Garth A},
  journal={Journal of chemical education},
  volume={99},
  number={10},
  pages={3406--3417},
  year={2022},
  publisher={ACS Publications}
}

@article{green2024understanding,
  title={Understanding Expressions of Self-Determination Theory in the Evaluation of IDEA-Themed VR Storytelling},
  author={Green, Kandice N and Yao, Shengjie and Lee, Heejae and Gratch, Lyndsay Michalik and Peters, David and Chock, T Makana},
  journal={Media and Communication},
  volume={12},
  year={2024}
}

@article{ijaz2020player,
  title={Player experience of needs satisfaction (PENS) in an immersive virtual reality exercise platform describes motivation and enjoyment},
  author={Ijaz, Kiran and Ahmadpour, Naseem and Wang, Yifan and Calvo, Rafael A},
  journal={International Journal of Human--Computer Interaction},
  volume={36},
  number={13},
  pages={1195--1204},
  year={2020},
  publisher={Taylor \& Francis}
}

@article{reer2022virtual,
  title={Virtual reality technology and game enjoyment: The contributions of natural mapping and need satisfaction},
  author={Reer, Felix and Wehden, Lars-Ole and Janzik, Robin and Tang, Wai Yen and Quandt, Thorsten},
  journal={Computers in Human Behavior},
  volume={132},
  pages={107242},
  year={2022},
  publisher={Elsevier}
}

@inproceedings{chen2024learning,
  title={Learning agent-based modeling with llm companions: Experiences of novices and experts using chatgpt \& netlogo chat},
  author={Chen, John and Lu, Xi and Du, Yuzhou and Rejtig, Michael and Bagley, Ruth and Horn, Mike and Wilensky, Uri},
  booktitle={Proceedings of the 2024 CHI Conference on Human Factors in Computing Systems},
  pages={1--18},
  year={2024}
}

@article{chiu2024teacher,
  title={Teacher support and student motivation to learn with Artificial Intelligence (AI) based chatbot},
  author={Chiu, Thomas KF and Moorhouse, Benjamin Luke and Chai, Ching Sing and Ismailov, Murod},
  journal={Interactive Learning Environments},
  volume={32},
  number={7},
  pages={3240--3256},
  year={2024},
  publisher={Taylor \& Francis}
}

@article{hew2023using,
  title={Using chatbots to support student goal setting and social presence in fully online activities: Learner engagement and perceptions},
  author={Hew, Khe Foon and Huang, Weijiao and Du, Jiahui and Jia, Chengyuan},
  journal={Journal of Computing in Higher Education},
  volume={35},
  number={1},
  pages={40--68},
  year={2023},
  publisher={Springer}
}

@inproceedings{jeong2025conversation,
  title={Conversation Progress Guide: UI system for enhancing self-efficacy in conversational AI},
  author={Jeong, Daeun and Shin, Sungbok and Jeong, Jongwook},
  booktitle={Proceedings of the 2025 CHI Conference on Human Factors in Computing Systems},
  pages={1--11},
  year={2025}
}

@inproceedings{kirchner2024outplay,
  title={Outplay your weaker self: A mixed-methods study on gamification to overcome procrastination in academia},
  author={Kirchner-Krath, Jeanine and Schmidt-Kraepelin, Manuel and Sch{\"o}bel, Sofia and Ullrich, Mathias and Sunyaev, Ali and Von Korflesch, Harald FO},
  booktitle={Proceedings of the 2024 CHI Conference on Human Factors in Computing Systems},
  pages={1--19},
  year={2024}
}

@article{stefanou2004supporting,
  title={Supporting autonomy in the classroom: Ways teachers encourage student decision making and ownership},
  author={Stefanou, Candice R and Perencevich, Kathleen C and DiCintio, Matthew and Turner, Julianne C},
  journal={Educational psychologist},
  volume={39},
  number={2},
  pages={97--110},
  year={2004},
  publisher={Taylor \& Francis}
}

@article{yin2024using,
  title={Using Educational Chatbots with Metacognitive Feedback to Improve Science Learning},
  author={Yin, Jiaqi and Zhu, Yi and Goh, Tiong-Thye and Wu, Wen and Hu, Yi},
  journal={Applied Sciences},
  volume={14},
  number={20},
  pages={9345},
  year={2024},
  publisher={MDPI}
}

@article{yin2025effects,
  title={Effects of different AI-driven Chatbot feedback on learning outcomes and brain activity},
  author={Yin, Jiaqi and Xu, Haoxin and Pan, Yafeng and Hu, Yi},
  journal={npj Science of Learning},
  volume={10},
  number={1},
  pages={17},
  year={2025},
  publisher={Nature Publishing Group UK London}
}

@inproceedings{cabales2019muse,
  title={Muse: Scaffolding metacognitive reflection in design-based research},
  author={Cabales, Victoria},
  booktitle={Extended Abstracts of the 2019 CHI Conference on Human Factors in Computing Systems},
  pages={1--6},
  year={2019}
}

@article{jang2010engaging,
  title={Engaging students in learning activities: It is not autonomy support or structure but autonomy support and structure.},
  author={Jang, Hyungshim and Reeve, Johnmarshall and Deci, Edward L},
  journal={Journal of educational psychology},
  volume={102},
  number={3},
  pages={588},
  year={2010},
  publisher={American Psychological Association}
}

@article{patall2008effects,
  title={The effects of choice on intrinsic motivation and related outcomes: a meta-analysis of research findings.},
  author={Patall, Erika A and Cooper, Harris and Robinson, Jorgianne Civey},
  journal={Psychological bulletin},
  volume={134},
  number={2},
  pages={270},
  year={2008},
  publisher={American Psychological Association}
}

@article{huang2023effects,
  title={Effects of artificial Intelligence--Enabled personalized recommendations on learners’ learning engagement, motivation, and outcomes in a flipped classroom},
  author={Huang, Anna YQ and Lu, Owen HT and Yang, Stephen JH},
  journal={Computers \& Education},
  volume={194},
  pages={104684},
  year={2023},
  publisher={Elsevier}
}

@article{white1959motivation,
  title={Motivation reconsidered: the concept of competence.},
  author={White, Robert W},
  journal={Psychological review},
  volume={66},
  number={5},
  pages={297},
  year={1959},
  publisher={American Psychological Association}
}

@article{multon1991relation,
  title={Relation of self-efficacy beliefs to academic outcomes: A meta-analytic investigation.},
  author={Multon, Karen D and Brown, Steven D and Lent, Robert W},
  journal={Journal of counseling psychology},
  volume={38},
  number={1},
  pages={30},
  year={1991},
  publisher={American Psychological Association}
}

@article{zhai2024effects,
  title={The effects of over-reliance on AI dialogue systems on students' cognitive abilities: a systematic review},
  author={Zhai, Chunpeng and Wibowo, Santoso and Li, Lily D},
  journal={Smart Learning Environments},
  volume={11},
  number={1},
  pages={28},
  year={2024},
  publisher={Springer}
}

@article{du2024transforming,
  title={Transforming language education: A systematic review of AI-powered chatbots for English as a foreign language speaking practice},
  author={Du, Jinming and Daniel, Ben Kei},
  journal={Computers and Education: Artificial Intelligence},
  volume={6},
  pages={100230},
  year={2024},
  publisher={Elsevier}
}

@inproceedings{zha2025mentigo,
  title={Mentigo: An Intelligent Agent for Mentoring Students in the Creative Problem Solving Process},
  author={Zha, Siyu and Liu, Yujia and Zheng, Chengbo and Xu, Jiaqi and Yu, Fuze and Gong, Jiangtao and Xu, Yingqing},
  booktitle={Proceedings of the 2025 CHI Conference on Human Factors in Computing Systems},
  pages={1--22},
  year={2025}
}

@article{azevedo2022lessons,
  title={Lessons learned and future directions of MetaTutor: Leveraging multichannel data to scaffold self-regulated learning with an intelligent tutoring system},
  author={Azevedo, Roger and Bouchet, Fran{\c{c}}ois and Duffy, Melissa and Harley, Jason and Taub, Michelle and Trevors, Gregory and Cloude, Elizabeth and Dever, Daryn and Wiedbusch, Megan and Wortha, Franz and others},
  journal={Frontiers in Psychology},
  volume={13},
  pages={813632},
  year={2022},
  publisher={Frontiers Media SA}
}

@article{neufeld2019exploring,
  title={Exploring the relationship between medical student basic psychological need satisfaction, resilience, and well-being: a quantitative study},
  author={Neufeld, Adam and Malin, Greg},
  journal={BMC medical education},
  volume={19},
  number={1},
  pages={405},
  year={2019},
  publisher={Springer}
}

@article{niemiec2009autonomy,
  title={Autonomy, competence, and relatedness in the classroom: Applying self-determination theory to educational practice},
  author={Niemiec, Christopher P and Ryan, Richard M},
  journal={Theory and research in Education},
  volume={7},
  number={2},
  pages={133--144},
  year={2009},
  publisher={Sage publications Sage UK: London, England}
}

@book{deci2013intrinsic,
  title={Intrinsic motivation and self-determination in human behavior},
  author={Deci, Edward L and Ryan, Richard M},
  year={2013},
  publisher={Springer Science \& Business Media}
}

@article{ryan2000self,
  title={Self-determination theory and the facilitation of intrinsic motivation, social development, and well-being.},
  author={Ryan, Richard M and Deci, Edward L},
  journal={American psychologist},
  volume={55},
  number={1},
  pages={68},
  year={2000},
  publisher={American Psychological Association}
}

@article{deci2000and,
  title={The" what" and" why" of goal pursuits: Human needs and the self-determination of behavior},
  author={Deci, Edward L and Ryan, Richard M},
  journal={Psychological inquiry},
  volume={11},
  number={4},
  pages={227--268},
  year={2000},
  publisher={Taylor \& Francis}
}

@article{lo2024influence,
  title={The influence of ChatGPT on student engagement: A systematic review and future research agenda},
  author={Lo, Chung Kwan and Hew, Khe Foon and Jong, Morris Siu-yung},
  journal={Computers \& Education},
  volume={219},
  pages={105100},
  year={2024},
  publisher={Elsevier}
}

@misc{wang2025effect,
  title={The effect of ChatGPT on students’ learning performance, learning perception, and higher-order thinking: Insights from a meta-analysis. Humanities and Social Sciences Communications, 11 (1), Article 59},
  author={Wang, J and Fan, W},
  year={2025}
}

@article{wu2024ai,
  title={Do AI chatbots improve students learning outcomes? Evidence from a meta-analysis},
  author={Wu, Rong and Yu, Zhonggen},
  journal={British Journal of Educational Technology},
  volume={55},
  number={1},
  pages={10--33},
  year={2024},
  publisher={Wiley Online Library}
}

@article{labadze2023role,
  title={Role of AI chatbots in education: systematic literature review},
  author={Labadze, Lasha and Grigolia, Maya and Machaidze, Lela},
  journal={International journal of Educational Technology in Higher education},
  volume={20},
  number={1},
  pages={56},
  year={2023},
  publisher={Springer}
}

@article{winget2022practical,
  title={A practical review of mastery learning},
  author={Winget, Marshall and Persky, Adam M},
  journal={American journal of pharmaceutical education},
  volume={86},
  number={10},
  pages={ajpe8906},
  year={2022},
  publisher={Elsevier}
}

@inproceedings{weinman2021improving,
  title={Improving instruction of programming patterns with faded parsons problems},
  author={Weinman, Nathaniel and Fox, Armando and Hearst, Marti A},
  booktitle={Proceedings of the 2021 chi conference on human factors in computing systems},
  pages={1--4},
  year={2021}
}

@article{atkinson2003transitioning,
  title={Transitioning from studying examples to solving problems: Effects of self-explanation prompts and fading worked-out steps.},
  author={Atkinson, Robert K and Renkl, Alexander and Merrill, Mary Margaret},
  journal={Journal of educational psychology},
  volume={95},
  number={4},
  pages={774},
  year={2003},
  publisher={American Psychological Association}
}

@article{vanlehn2011relative,
  title={The relative effectiveness of human tutoring, intelligent tutoring systems, and other tutoring systems},
  author={VanLehn, Kurt},
  journal={Educational psychologist},
  volume={46},
  number={4},
  pages={197--221},
  year={2011},
  publisher={Taylor \& Francis}
}

@inproceedings{kazemitabaar2024codeaid,
  title={Codeaid: Evaluating a classroom deployment of an llm-based programming assistant that balances student and educator needs},
  author={Kazemitabaar, Majeed and Ye, Runlong and Wang, Xiaoning and Henley, Austin Zachary and Denny, Paul and Craig, Michelle and Grossman, Tovi},
  booktitle={Proceedings of the 2024 chi conference on human factors in computing systems},
  pages={1--20},
  year={2024}
}

@inproceedings{lieb2024student,
  title={Student interaction with newtbot: An llm-as-tutor chatbot for secondary physics education},
  author={Lieb, Anna and Goel, Toshali},
  booktitle={Extended Abstracts of the CHI Conference on Human Factors in Computing Systems},
  pages={1--8},
  year={2024}
}

@article{knievel2025aitee,
  title={AITEE--Agentic Tutor for Electrical Engineering},
  author={Knievel, Christopher and Bernhardt, Alexander and Bernhardt, Christian},
  journal={arXiv preprint arXiv:2505.21582},
  year={2025}
}

\appendix
\appendix
\section{Appendix}\label{sec:appendix}

\subsection{Demographics Survey Questionnaire}\label{subsec:survey}
  \begin{enumerate}
    \item [\textbf{[A1]}] What is your age?
    
    \item [\textbf{[A2]}] What is your field of study? 
    \begin{enumerate}
    \item[A.] Aeronautics \& Astronautics
    \item[B.] Bioengineering
    \item[C.] Chemical Engineering
    \item[D.] Civil \& Environmental Engineering 
    \item[E.] Computer Science \& Engineering 
    \item[F.] Electrical \& Computer Engineering 
    \item[G.] Human Centered Design \& Engineering 
    \item[H.] Industrial \& Systems Engineering 
    \item[I.] Materials Science \& Engineering 
    \item[J.] Mechanical Engineering
    \item[K.] Physics, Applied Physics, Chemistry, or other Physical Science
    \item[L.] Biology, or other Natural Science 
    \item[M.] Other (please specify): \rule{3cm}{0.4pt}
    \end{enumerate}
    
    \item [\textbf{[A3]}] What is your current academic standing? 
    \begin{enumerate}
    \item[A.] Freshman
    \item[B.] Sophomore
    \item[C.] Junior
    \item[D.] Senior
    \item[E.] Alumni (Graduated)
    \item[F.] Other (please specify): \rule{3cm}{0.4pt}
    \end{enumerate}
    
    \item [\textbf{[A4]}] What gender do you identify as? 
    \begin{enumerate}
    \item[A.] Female
    \item[B.] Male
    \item[C.] Non-binary
    \item[D.] Transgender
    \item[E.] Prefer not to answer
    \item[F.] Other (please specify): \rule{3cm}{0.4pt}
    \end{enumerate}

    \item [\textbf{[A5]}] What is your race? (Select all that apply)
    \begin{enumerate}
      \item[A.] White
      \item[B.] Black or African American
      \item[C.] American Indian or Alaska Native
      \item[D.] Asian
      \item[E.] Native Hawaiian or Other Pacific Islander
      \item[F.] Other (please specify): \rule{3cm}{0.4pt}
    \end{enumerate}

    \item [\textbf{[A6]}] Are you Hispanic or Latino or Spanish origin?
    \begin{enumerate}
      \item[A.] Yes
      \item[B.] No
    \end{enumerate}
    
    \item [\textbf{[A7]}] What is your marital status? 
    \begin{enumerate}
      \item[A.] Single (never married)
      \item[B.] Married or in a domestic partnership
      \item[C.] Widowed, divorced, or separated
    \end{enumerate}
    \end{enumerate}

\subsection{Attention Measures}\label{subsec:attention}
In general, please rate how often each statement applies to you on a scale from 0 (Never) to 4 (Very often).
    \begin{table}[!ht]
    \small
    \caption{Likert-scale Items for Attention.}
    \centering
\resizebox{\columnwidth}{!}{
    \begin{tabular}{p{1cm} p{13.5cm}}
        \toprule
        \textbf{Item} & \textbf{Question} \\
        \midrule
        IA1 & How often do you make careless mistakes when you have to work on a boring or difficult project? \\
        IA2 & How often do you have difficulty keeping your attention when you are doing boring or repetitive work? \\
        IA3 & How often do you have difficulty concentrating on what people say to you, even when they are speaking to you directly? \\
        IA4 & How often do you have trouble wrapping up the fine details of a project, once the challenging parts have been done? \\
        IA5 & How often do you have difficulty getting things in order when you have to do a task that requires organization? \\
        IA6 & When you have a task that requires a lot of thought, how often do you avoid or delay getting started? \\
        IA7 & How often do you misplace or have difficulty finding things at home or at work? \\
        IA8 & How often are you distracted by activity or noise around you? \\
        IA9 & How often do you have problems remembering appointments or obligations? \\
        \bottomrule
    \end{tabular}
    }    \label{tab:c1_items}
    \end{table}

\subsection{Baseline Measures}\label{subsec:baseline}
For the following questions (Table \ref{tab:autonomy_competence_items}), please respond with respect to your general experiences in classes in your major/discipline (1: Strongly Disagree; 2: Disagree; 3: Neutral; 4: Agree; 5: Strongly Agree).

\subsection{Perceived Measures}\label{subsec:preceived}
For the following questions (Table \ref{tab:autonomy_competence_items_ai}), an AI tool refers to software specifically built with artificial intelligence techniques and characterized by a personalized or customized approach to learning that is used by a student (whether recommended by an instructor or not) to support learning and performance in the student’s courses.
\begin{enumerate}
 \item [\textbf{[D1]}] Which of the following AI-driven learning tools have you used most frequently to support your coursework? \textsuperscript{*}
    \begin{enumerate}
      \item[A.] Adaptive learning platforms. For example, Pearson’s Mastering Engineering offers real-time feedback on solving problems—such as identifying where you went wrong and providing hints on homework assignments)
      \item[B.] AI-powered tutoring. For instance, Q Chat offered in Quizlet transforms your Quizlet study set into an interactive dialogue—quizzing you, spotting mistakes, and tailoring follow up prompts in real time.
      \item[C.] Chatbots (e.g., ChatGPT, Gemini, Claude, Microsoft Copilot)
      \item[D.] Virtual Reality (VR) or Augmented Reality (AR) learning applications. For example, Physics 122 at a large public research university uses VR to create a virtual environment that helps students analyze electric forces, observe interactions between electrons, and explore the properties of unknown particles through simulated experiments.
      \item[E.] Other (please specify): \rule{3cm}{0.4pt}
    \end{enumerate} 
\end{enumerate} 
In the previous question, you responded that among all the AI tools available, you used \textless AI tool from \textbf{[D1]}\textgreater{} most frequently to support you in your learning and academic performance. Please answer the following questions with respect to \textless AI tool from \textbf{[D1]}\textgreater{}, hereafter referred to as ``this AI tool.''

 \begin{table}[!ht]
        \small
        \caption{Likert-scale Items for Baseline Measures.}
        \centering
        \resizebox{\columnwidth}{!}{
        \begin{tabular}{p{3cm} p{13cm}}
        \toprule
        \textbf{Constructs} & \textbf{Items} \\
        \midrule
        \textbf{Autonomy (AU)}
        & 1.\;I feel a sense of choice and freedom in the curriculum I undertake. \\
        & 2.\;I feel that the decisions I make in my education truly reflect what I want to do. \\
        & 3.\;Most of the tasks I complete in my curriculum feel like requirements I have to fulfill rather than choices I make. \\
        & 4.\;I feel that, in my coursework and other activities, I have been doing what really interests me. \\
        \midrule
        \textbf{Competence (CO)}
        & 1.\;I feel confident that I can successfully complete challenging tasks. \\
        & 2.\;I often have serious doubts about whether I can excel. \\
        & 3.\;I sometimes feel disappointed with my overall performance. \\
        & 4.\;I feel insecure about my abilities in my coursework. \\
        & 5.\;When working on coursework, I sometimes feel like a failure because of the mistakes I make. \\
        \midrule
        \textbf{Relatedness (RE)}
        & 1.\;I feel that the classmates and instructors I care about also care about me. \\
        & 2.\;I feel connected with others who are important to me. \\
        & 3.\;I feel that people who are important to me are cold or distant toward me. \\
        & 4.\;I feel that I am accepted by my peers. \\
        \midrule
        \textbf{Self-Regulated Learning (SR)}
        & 1.\;I feel comfortable connecting new knowledge with what I already know in my coursework. \\
        & 2.\;I feel confident that I can develop my own ideas when materials are given in my coursework. \\
        & 3.\;I always consider how to approach a problem before beginning to solve it. \\
        & 4.\;If I struggle with coursework, I know where to find help. \\
        & 5.\;I feel comfortable taking advantage of external feedback (e.g., advice from professors or TAs) to improve in coursework. \\
        \midrule
        \textbf{Self-Efficacy (SE)}
        & 1.\;I can remain calm when facing difficulties because I trust my abilities. \\
        & 2.\;When I am confronted with a challenge in my courses, I can usually find multiple solutions. \\
        & 3.\;Whatever comes my way, I can usually handle it. \\
        & 4.\;I feel prepared for most of the demands in my coursework and academic program. \\
        \midrule
        \textbf{IMC 1}
        & I feel that paying close attention to instructions is important for accurately completing this survey. Please select “Agree” for this statement. \\
        \bottomrule
        \end{tabular}
        }
        \label{tab:autonomy_competence_items}
    \end{table}

\begin{table}[!ht]
        \small
        \caption{Likert-scale Items for Perceived Measures.}
        \centering
        \resizebox{\columnwidth}{!}{
        \begin{tabular}{p{3cm} p{13cm}}
        \toprule
        \textbf{Constructs} & \textbf{Items} \\
        \midrule
        \textbf{Autonomy (PAU)}
        & 1.\; \ldots\ has improved my sense of choice and freedom in the curriculum I undertake. \\
        & 2.\; \ldots\ has helped me to make decisions that truly reflect what I want to do. \\
        & 3.\; \ldots\ has made me feel more like the tasks I have to complete are by choice rather than simply required as part of my degree. \\
        & 4.\; \ldots\ has made me feel that, in my coursework and other activities, I have been doing what really interests me. \\
        \midrule
        \textbf{Competence (PCO)}
        & 1.\; \ldots\ has caused me to feel more confident that I can successfully complete challenging tasks. \\
        & 2.\; \ldots\ has made me have less serious doubts about whether I can excel. \\
        & 3.\; \ldots\ has reduced my disappointment with my overall performance. \\
        & 4.\; \ldots\ has made me feel less insecure about my abilities in my coursework. \\
        & 5.\; \ldots\ has made me feel less like a failure when I make mistakes in working on coursework. \\
        \midrule
        \textbf{Relatedness (PRE)}
        & 1.\; \ldots\ has made me feel that the classmates and instructors I care about also care about me. \\
        & 2.\; \ldots\ has made me feel more connected with others who are important to me. \\
        & 3.\; \ldots\ has made me feel that people who are important to me are less cold or distant toward me. \\
        & 4.\; \ldots\ has made me feel that I am more accepted by my peers. \\
        \midrule
        \textbf{IMC 2}
        & \ldots\ has enabled me to complete every assignment before it is released by my instructor. \\
        \bottomrule
        \end{tabular}
        }
        \label{tab:autonomy_competence_items_ai}
    \end{table}

\end{document}